\documentclass{nature}

\usepackage[british]{babel}
\usepackage[utf8]{inputenc}
\usepackage{babelbib}
\usepackage{url}
\usepackage{graphicx}
\usepackage{comment}
\usepackage{subfig}
\usepackage{calc}
\usepackage{floatflt}
\usepackage{amssymb, amsmath, amsthm}
\usepackage{tabularx}
\usepackage{multirow}
\usepackage{array}
\usepackage{xcolor}
\usepackage{bm}
\usepackage{stmaryrd}
\usepackage{stackrel}
\usepackage{algpseudocode}
\usepackage{algorithm}
\usepackage{rotating}
\usepackage{mwe,tikz}\usepackage[percent]{overpic}
\usepackage{float}
\usepackage{listings}
\usepackage[colorlinks=true,allcolors=blue]{hyperref}

\usepackage{numprint}
\usepackage{memhfixc}
\usepackage{makecell}
\usepackage{siunitx}
\usepackage{textcomp}
\usepackage{arydshln}

\usepackage{amsmath}

\usepackage{mathtools}

\title{Neural 360$^\circ$ Structured Light with Learned Metasurfaces}

\author{Eunsue Choi$^{1*}$, Gyeongtae Kim$^{2*}$,  Jooyeong Yun$^{2}$, Yujin Jeon$^{1}$, Junsuk Rho$^{2,3,4,5\dagger}$, Seung-Hwan Baek$^{1\dagger}$}

\usepackage{lipsum}

\begin{document}

\maketitle



\definecolor{brightray}{rgb}{0.8,0.8,0.8}
\definecolor{Gray}{rgb}{0.5,0.5,0.5}
\definecolor{darkblue}{rgb}{0,0,0.7}
\definecolor{orange}{rgb}{1,.5,0} 
\definecolor{red}{rgb}{1,0,0} 
\definecolor{blue}{rgb}{0,0,1} 
\definecolor{darkgreen}{rgb}{0,0.7,0} 
\definecolor{darkred}{rgb}{0.7,0,0} 

\newcommand{\heading}[1]{\noindent\textbf{#1}}
\newcommand{\note}[1]{{{\textcolor{orange}{#1}}}}
\newcommand{\todo}[1]{{\textcolor{red}{TODO: #1}}}
\newcommand{\changed}[1]{{\textcolor{blue}{#1}}}
\newcommand{\removed}[1]{{\textcolor{brightray}{{#1}}}}
\newcommand{\revision}[1]{{{#1}}}
\newcommand{\place}[1]{ \begin{itemize}\item\textcolor{darkblue}{#1}\end{itemize}}
\newcommand{\de}{\mathrm{d}}

\newcommand{\BEAS}{\begin{eqnarray*}}
\newcommand{\EEAS}{\end{eqnarray*}}
\newcommand{\BEA}{\begin{eqnarray}}
\newcommand{\EEA}{\end{eqnarray}}
\newcommand{\BEQ}{\begin{equation}}
\newcommand{\EEQ}{\end{equation}}
\newcommand{\BIT}{\begin{itemize}}
\newcommand{\EIT}{\end{itemize}}
\newcommand{\BNUM}{\begin{enumerate}}
\newcommand{\ENUM}{\end{enumerate}}

\newcommand{\BA}{\begin{array}}
\newcommand{\EA}{\end{array}}

\newcommand{\eg}{{\it e.g.}}
\newcommand{\ie}{{\it i.e.}}
\newcommand{\etc}{{\it etc.}}

\newcommand{\ones}{\mathbf 1}

\newcommand{\reals}{{\mbox{\bf R}}}
\newcommand{\integers}{{\mbox{\bf Z}}}
\newcommand{\eqbydef}{\mathrel{\stackrel{\Delta}{=}}}
\newcommand{\complex}{{\mbox{\bf C}}}
\newcommand{\symm}{{\mbox{\bf S}}}  

\newcommand{\Span}{\mbox{\textrm{span}}}
\newcommand{\Range}{\mbox{\textrm{range}}}
\newcommand{\nullspace}{{\mathcal N}}
\newcommand{\range}{{\mathcal R}}
\newcommand{\Nullspace}{\mbox{\textrm{nullspace}}}
\newcommand{\Rank}{\mathop{\bf Rank}}
\newcommand{\Tr}{\mathop{\bf Tr}}
\newcommand{\diag}{\mathop{\bf diag}}
\newcommand{\lambdamax}{{\lambda_{\rm max}}}
\newcommand{\lambdamin}{\lambda_{\rm min}}

\newcommand{\Expect}{\mathop{\bf E{}}}
\newcommand{\Prob}{\mathop{\bf Prob}}
\newcommand{\erf}{\mathop{\bf erf}}

\newcommand{\Co}{{\mathop {\bf Co}}}
\newcommand{\co}{{\mathop {\bf Co}}}
\newcommand{\dist}{\mathop{\bf dist{}}}
\newcommand{\Ltwo}{{\bf L}_2}
\newcommand{\QED}{~~\rule[-1pt]{8pt}{8pt}}\def\qed{\QED}
\newcommand{\approxleq}{\mathrel{\smash{\makebox[0pt][l]{\raisebox{-3.4pt}{\small$\sim$}}}{\raisebox{1.1pt}{$<$}}}}
\newcommand{\epi}{\mathop{\bf epi}}

\newcommand{\vol}{\mathop{\bf vol}}
\newcommand{\Vol}{\mathop{\bf vol}}
\newcommand{\Card}{\mathop{\bf card}}

\newcommand{\dom}{\mathop{\bf dom}}
\newcommand{\aff}{\mathop{\bf aff}}
\newcommand{\cl}{\mathop{\bf cl}}
\newcommand{\Angle}{\mathop{\bf angle}}
\newcommand{\intr}{\mathop{\bf int}}
\newcommand{\relint}{\mathop{\bf rel int}}
\newcommand{\bd}{\mathop{\bf bd}}
\newcommand{\vect}{\mathop{\bf vec}}
\newcommand{\dsp}{\displaystyle}
\newcommand{\foequal}{\simeq}
\newcommand{\VOL}{{\mbox{\bf vol}}}
\newcommand{\xopt}{x^{\rm opt}}

\newcommand{\Xb}{{\mbox{\bf X}}}
\newcommand{\xst}{x^\star}
\newcommand{\varphist}{\varphi^\star}
\newcommand{\lambdast}{\lambda^\star}
\newcommand{\Zst}{Z^\star}
\newcommand{\fstar}{f^\star}
\newcommand{\xstar}{x^\star}
\newcommand{\xc}{x^\star}
\newcommand{\lambdac}{\lambda^\star}
\newcommand{\lambdaopt}{\lambda^{\rm opt}}

\newcommand{\geqK}{\mathrel{\succeq_K}}
\newcommand{\gK}{\mathrel{\succ_K}}
\newcommand{\leqK}{\mathrel{\preceq_K}}
\newcommand{\lK}{\mathrel{\prec_K}}
\newcommand{\geqKst}{\mathrel{\succeq_{K^*}}}
\newcommand{\gKst}{\mathrel{\succ_{K^*}}}
\newcommand{\leqKst}{\mathrel{\preceq_{K^*}}}
\newcommand{\lKst}{\mathrel{\prec_{K^*}}}
\newcommand{\geqL}{\mathrel{\succeq_L}}
\newcommand{\gL}{\mathrel{\succ_L}}
\newcommand{\leqL}{\mathrel{\preceq_L}}
\newcommand{\lL}{\mathrel{\prec_L}}
\newcommand{\geqLst}{\mathrel{\succeq_{L^*}}}
\newcommand{\gLst}{\mathrel{\succ_{L^*}}}
\newcommand{\leqLst}{\mathrel{\preceq_{L^*}}}
\newcommand{\lLst}{\mathrel{\prec_{L^*}}}

\newtheorem{theorem}{Theorem}[section]
\newtheorem{corollary}{Corollary}[theorem]
\newtheorem{lemma}[theorem]{Lemma}
\newtheorem{proposition}[theorem]{Proposition}

\newenvironment{algdesc}%
{\begin{quote}}{\end{quote}}

\def\figbox#1{\framebox[\hsize]{\hfil\parbox{0.9\hsize}{#1}}}

\makeatletter
\long\def\@makecaption#1#2{
   \vskip 9pt
   \begin{small}
   \setbox\@tempboxa\hbox{{\bf #1:} #2}
   \ifdim \wd\@tempboxa > 5.5in
        \begin{center}
        \begin{minipage}[t]{5.5in}
        \addtolength{\baselineskip}{-0.95pt}
        {\bf #1:} #2 \par
        \addtolength{\baselineskip}{0.95pt}
        \end{minipage}
        \end{center}
   \else
    \hbox to\hsize{\hfil\box\@tempboxa\hfil}
   \fi
   \end{small}\par
}
\makeatother

\newcounter{oursection}
\newcommand{\oursection}[1]{
 \addtocounter{oursection}{1}
 \setcounter{equation}{0}
 \clearpage \begin{center} {\Huge\bfseries #1} \end{center}
 {\vspace*{0.15cm} \hrule height.3mm} \bigskip
 \addcontentsline{toc}{section}{#1}
}
\newcommand{\oursectionf}[1]{  
 \addtocounter{oursection}{1}
 \setcounter{equation}{0}
 \foilhead[-.5cm]{#1 \vspace*{0.8cm} \hrule height.3mm }
 \LogoOn
}
\newcommand{\oursectionfl}[1]{  
 \addtocounter{oursection}{1}
 \setcounter{equation}{0}
 \foilhead[-1.0cm]{#1}
 \LogoOn
}

\newcommand{\Mat}[1]    {{\ensuremath{\mathbf{\uppercase{#1}}}}} 
\newcommand{\Vect}[1]   {{\ensuremath{\mathbf{\lowercase{#1}}}}} 
\newcommand{\Vari}[1]   {{\ensuremath{\mathbf{\lowercase{#1}}}}} 
\newcommand{\Id}				{\mathbb{I}} 
\newcommand{\Diag}[1] 	{\operatorname{diag}\left({ #1 }\right)} 
\newcommand{\Opt}[1] 	  {{#1}_{\text{opt}}} 
\newcommand{\CC}[1]			{{#1}^{*}} 
\newcommand{\Op}[1]     {\Mat{#1}} 
\newcommand{\mini}[1] {{\mbox{argmin}}_{#1} \: \: } 
\newcommand{\argmin}[1] {\underset{{#1}}{\mathop{\rm argmin}} \: \: } 
\newcommand{\argmax}[1] {\underset{{#1}}{\mathop{\rm argmax}} \: \: } 
\newcommand{\minimize}{\mathop{\rm minimize} \: \:}
\newcommand{\minimizeu}[1]{\underset{{#1}}{\mathop{\rm minimize}} \: }
\newcommand{\grad}      {\nabla}
\newcommand{\kron}{\otimes} 

\newcommand{\gradt}     {\grad_\z}
\newcommand{\gradx}     {\grad_\x}
\newcommand{\Drv}     	{\Mat{D}} 
\newcommand{\step}      {\text{\textbf{step}}}
\newcommand{\prox}[1]   {\mathbf{prox}_{#1}}
\newcommand{\ind}[1]    {\operatorname{ind}_{#1}}
\newcommand{\proj}[1]   {\Pi_{#1}}
\newcommand{\pointmult}{\odot} 
\newcommand{\rr}   {\mathcal{R}}

\newcommand{\Basis}{\Mat{D}}         		
\newcommand{\Corr}{\Mat{C}}             
\newcommand{\conv}{\ast} 
\newcommand{\meas}{\Vect{b}}            
\newcommand{\Img}{I}                    
\newcommand{\img}{\Vect{i}}             
\newcommand{\vv}{\Vect{v}}
\newcommand{\p}{\Vect{p}}
\newcommand{\Splitvar}{T}                
\newcommand{\splitvar}{\Vect{t}}         
\newcommand{\Splitbasis}{J}                
\newcommand{\splitbasis}{\Vect{j}}         
\newcommand{\var}{\Vari{z}}

\newcommand{\FT}[1]			{\mathcal{F}\left( {#1} \right)} 
\newcommand{\IFT}[1]			{\mathcal{F}^{-1}\left( {#1} \right)} 

\newcommand{\func}{f}
\newcommand{\fMat}{\Mat{K}}

\newcommand{\avar}{\Vari{v}}
\newcommand{\aspvar}{\Vari{z}}

\newcommand{\mask}{\Mat{M}}

\newcommand{\Pen}      		{F} 
\newcommand{\cardset}     {\mathcal{C}}
\newcommand{\Dat}      		{G} 
\newcommand{\Reg}      		{\Gamma} 

\newcommand{\Trans}{\mathbf{\uppercase{T}}} 
\newcommand{\Ph}{\mathbf{\uppercase{\Phi}}} 

\newcommand{\Tvec}{\Vect{T}} 
\newcommand{\Bvec}{\Vect{B}} 

\newcommand{\Wt}{\Mat{W}} 

\newcommand{\Perm}{\Mat{P}} 
\newcommand{\Cblur}{\Mat{C}} 

\newcommand{\DiagFactor}[1]     {\Mat{O}_{ #1 }}  

\newcommand{\Proj}{\Mat{P}}             

\newcommand{\Vector}[1]{\mathbf{#1}}
\newcommand{\Matrix}[1]{\mathbf{#1}}
\newcommand{\Tensor}[1]{\boldsymbol{\mathscr{#1}}}
\newcommand{\TensorUF}[2]{\Matrix{#1}_{(#2)}}

\newcommand{\MatrixKP}[1]{\Matrix{#1}_{\otimes}}
\newcommand{\MatrixKPN}[2]{\Matrix{#1}_{\otimes}^{#2}}

\newcommand{\MatrixKRP}[1]{\Matrix{#1}_{\odot}}
\newcommand{\MatrixKRPN}[2]{\Matrix{#1}_{\odot}^{#2}}

\newcommand{\HP}{\circ}
\newcommand{\HD}{\oslash}

\newcommand{\leftDB}{\left[ \! \left[}
\newcommand{\rightDB}{\right] \! \right]}

\newcommand{\transpose}{T}

\newcommand*\sstrut[1]{\vrule width0pt height0pt depth#1\relax}

\newcommand{\inlineeqnum}{\refstepcounter{equation}~~\mbox{(\theequation)}}
\newcommand{\eqname}[1]{\tag*{#1~(\theequation)}\refstepcounter{equation}}

\newcommand{\lambdas}{\boldsymbol{\lambda}}
\newcommand{\alb}{\boldsymbol{\alpha}} 	
\newcommand{\depth}{\boldsymbol{z}} 	
\newcommand{\albi}{\alpha} 	
\newcommand{\depthi}{z} 	
\newcommand{\ambient}{s}
\newcommand{\jitter}{w}
\newcommand{\z}{\Vect{z}} 							
\newcommand{\x}{\Vect{x}}             	
\newcommand{\y}{\Vect{y}}             	
\newcommand{\Kvar}{\Mat{K}}
\newcommand{\lagrangemult}{\boldsymbol{\nu}}
\newcommand{\scaledlagrange}{\Vect{u}}
\newcommand{\eps}{\epsilon}
\newcommand{\vp}{\Vect{v}}

\begin{affiliations}
 \item Department of Computer Science and Engineering, Pohang University of Science and Technology (POSTECH), Pohang 37673, Republic of Korea
 \item Department of Mechanical Engineering, Pohang University of Science and Technology (POSTECH), Pohang 37673, Republic of Korea
 \item Department of Chemical Engineering, Pohang University of Science and Technology (POSTECH), Pohang 37673, Republic of Korea
 \item POSCO-POSTECH-RIST Convergence Research Center for Flat Optics and Metaphotonics, Pohang 37673, Republic of Korea
 \item National Institute of Nanomaterials Technology (NINT), Pohang 37673, Republic of Korea
  \item [$*$] Equal contribution
 \item [$\dagger$] Corresponding author. E-mail: shwbaek@postech.ac.kr, jsrho@postech.ac.kr
\end{affiliations}
\begin{abstract}
Structured light has proven instrumental in 3D imaging, LiDAR, and holographic light projection. Metasurfaces, comprised of sub-wavelength-sized nanostructures, facilitate 180$^\circ$ field-of-view (FoV) structured light, circumventing the restricted FoV inherent in traditional optics like diffractive optical elements. However, extant metasurface-facilitated structured light exhibits sub-optimal performance in downstream tasks, due to heuristic pattern designs such as periodic dots\cite{kim2022metasurface} that do not consider the objectives of the end application.
In this paper, we present neural 360$^\circ$ structured light, driven by learned metasurfaces. 
We propose a differentiable framework, that encompasses a computationally-efficient 180$^\circ$ wave propagation model and a task-specific reconstructor, and exploits both transmission and reflection channels of the metasurface. Leveraging a first-order optimizer within our differentiable framework, we optimize the metasurface design, thereby realizing neural 360$^\circ$ structured light.
We have utilized neural 360$^\circ$ structured light for holographic light projection and 3D imaging. 
Specifically, we demonstrate the first 360$^\circ$ light projection of complex patterns, enabled by our propagation model that can be computationally evaluated 50,000$\times$ faster than the Rayleigh-Sommerfeld propagation.
For 3D imaging, we improve depth-estimation accuracy by 5.09$\times$ in RMSE compared to the heuristically-designed structured light\cite{kim2022metasurface}.
Neural 360$^\circ$ structured light promises robust 360$^\circ$ imaging and display for robotics, extended-reality systems, and human-computer interactions.

\end{abstract}



The structured-light technique, designed to project a specific light pattern with wavefront modulation, has been applied in various fields including 3D imaging, holographic display, light detection and ranging, and augmented and virtual reality display\cite{geng2011structured}. Diffractive optical elements (DOEs) and spatial light modulators (SLMs), paired with laser illumination, have traditionally been employed to facilitate structured light\cite{DAMMANN1971312, Zhou:95,He:21}. Recent methods have leveraged computational end-to-end optimization methods to achieve high projection fidelity for a range of patterns, exemplified in SLM-based dynamic holographic displays\cite{shi2021towards,tseng2023neural,peng2020neural} and DOE-based 3D imaging\cite{baek2021polka}. 
Nonetheless, these methods face a fundamental constraint: a limited FoV due to restricted diffraction angles from the micron-sized structures of DOE and SLM. This limitation hinders the application of structured light for emerging 360$^\circ$ 3D imaging and display in augmented reality devices, robotics, autonomous vehicles, and human-computer interactions.

To tackle this problem, previous methods have incorporated mechanical elements, such as rotating heads or microelectromechanical systems\cite{yang2019optical, zhang2016wide}. However, such mechanical elements often introduce system complexity, increase power consumption, and are unsuitable for high-speed operation. Recently, metasurfaces have been recognized as a potential solution to overcome the FoV issue. Comprising ultra-thin arrays of subwavelength structures, metasurfaces can manipulate light at the nanoscale by altering multiple incident light properties including phase, amplitude and polarization\cite{arbabi2015dielectric,doi:10.1126/science.1210713,so2022revisiting,overvig2019dielectric,kim2021pixelated}. This capability allows precise control of light propagation and the generation of complex optical wavefronts, enabling functionalities such as beam steering\cite{doi:10.1126/science.1210713, wu2019dynamic, doi:10.1021/acsnano.1c08597}, focusing\cite{wang2018broadband,khorasaninejad2017metalenses,doi:10.1021/acsami.9b07774,yoon2020single}, and holography\cite{zheng2015metasurface,ni2013metasurface,huang2013three,yoon2018crypto,kim2020stimuli}. In particular, since their subwavelength nanostructures can achieve diffraction angles up to 90$^\circ$, metasurfaces enable 180$^\circ$ FoV structured light, eliminating the need for mechanical scanning.

To design 180$^\circ$ FoV structured light with metasurfaces, the intensity distribution of propagated light after the wavefront modulation should be easily calculated at target angles and distances and the phase map of the metasurface must be optimized for various downstream tasks. Here, it is important to use computationally-efficient wave-propagation models to apply algorithmic techniques in its optimization process. For example, Fast-Fourier transformation (FFT) is commonly used to efficiently compute Fresnel and Fraunhofer diffraction integral because their formulas are expressed as Fourier transform evaluated at spatial frequencies. However, these formulas are derived using paraxial approximation and hence the computed results are only valid for small range of diffraction angles\cite{huang2013three,wen2015helicity}. To accurately model wave-propagation over large range of diffracted region, the Rayleigh-Sommerfeld diffraction integral which calculates intensity at any target position by integrating the effect of every source position can be directly used\cite{heurtley1973scalar,totzeck1991validity} or used in parallel with Fraunhofer diffraction to compensate inaccurate modeling at large angle\cite{chen20202pi}. However, direct numerical integration of every source point is computationally intensive and hence makes it intractable to be in conjunction with algorithmic optimization techniques that involves multiple iterations and updates. Therefore, a wave-propagation model which can cover large range of angle while being computationally efficient is required. To overcome this challenge, several previous works have employed structural constraints on metasurfaces such as arranging meta-atoms in super-cells to exploit higher-order diffraction orders. Consequently, the large number of diffraction orders could be efficiently modeled over full-space\cite{li2018full,kim2022metasurface}, however these designs are limited to generating periodic or random array of dots for the structured-light patterns, often leading to sub-optimal performance on downstream tasks.

\begin{figure}[t]
	\centering
		\includegraphics[width=\columnwidth]{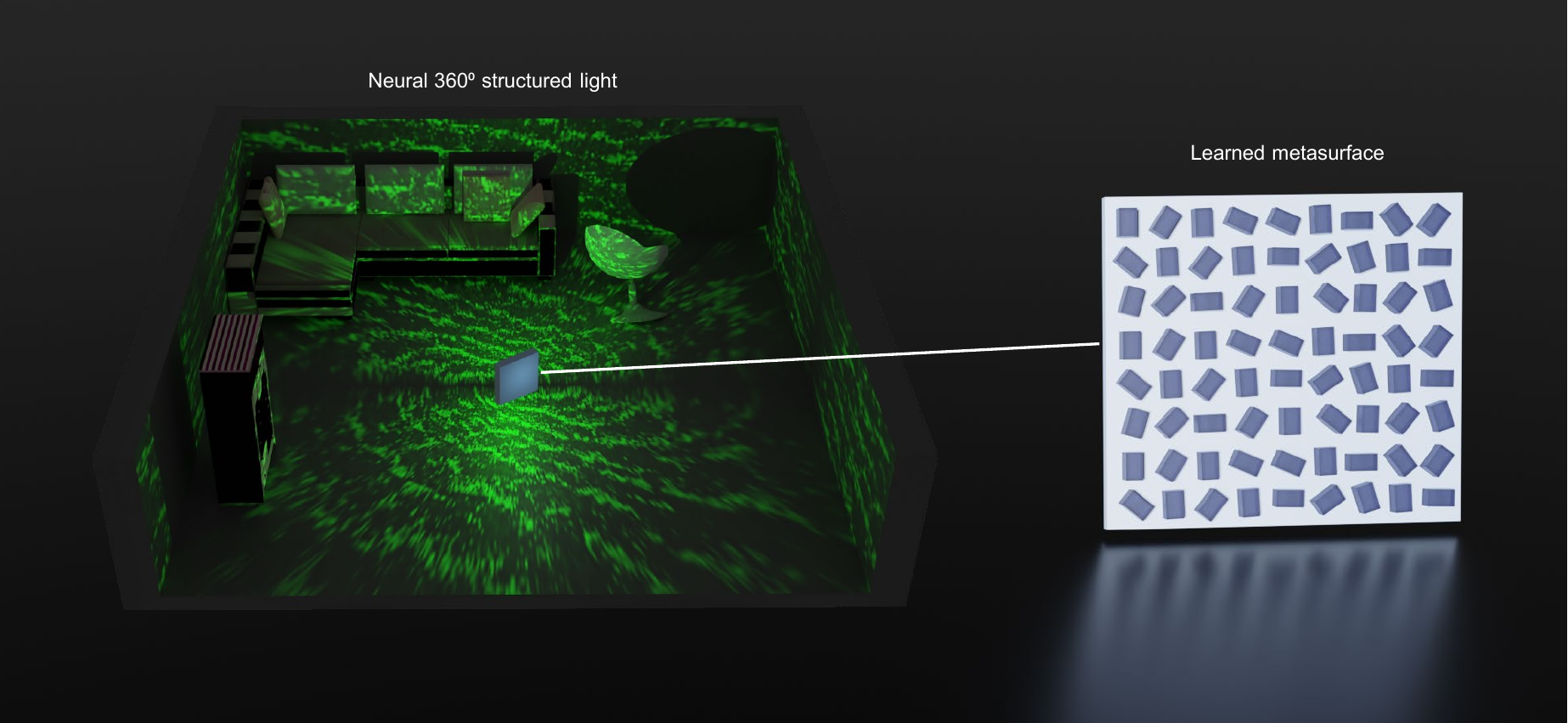}
		\caption{Neural 360$^\circ$ structured light enables 360$^\circ$ light projection of a computationally-designed pattern with a learned metasurface.}
		\label{fig:teaser}
\end{figure}

Here, we propose neural 360$^\circ$ structured light, enabled by a computationally-designed metasurface optimized for a downstream task. Neural 360$^\circ$ structured light allows for high-fidelity 360$^\circ$ light projection of intricate patterns without mechanical movement (Fig.~\ref{fig:teaser}). To this end, we present a differentiable framework consisting of a wave-propagation model that accurately models the 180$^\circ$ far-field propagation of metasurface-diffracted light in a differentiable manner, and a downstream-task-specific reconstructor. Leveraging the high efficiency of reflection and transmission channels of a metasurface and the differentiable 180$^\circ$ wave propagation 50,000$\times$ faster than the Rayleigh-Sommerfeld propagation, we learn the 360$^\circ$ structured light with the end loss of the downstream task. 
We apply neural 360$^\circ$ structured light for holographic light projection and 3D imaging.
For 3D imaging, we demonstrate 5.09$\times$ reduction of depth-reconstruction RMSE compared to the multi-dot full-space structured light\cite{kim2022metasurface}. 
For light projection, we show the first demonstration of 360$^\circ$ complex-pattern light projection. 

\section*{Results}
\subsection{Design of neural 360$^\circ$ structured light.}

Here, we present a differentiable and computationally-efficient wave propagation model capable of describing the propagation of light interacted with metasurfaces to the propagated 180$^\circ$ far-field wavefront.
We start by reframing the target coordinate of Rayleigh-Sommerfeld diffraction integral to the spherical coordinate as 
\begin{align}
\label{eq:sph_coord}
 U(\rho, \theta, \phi) = \frac{1}{\lambda j} \int\int U'(x', y')\frac{e^{jkr}}{r}\left(1-\frac{1}{jkr}\right)\frac{\rho\sin{\theta}\sin{\phi}}{r}\, dx'dy',
\end{align}
where $U'$ and $U$ refer to the source and target wavefronts defined in Cartesian coordinate and spherical coordinate, respectively.  
$\rho$ represents the distance between the coordinate origin $o$ and the target point $p_{t}$ while $r$ denotes the distance between the source point $p_{m}$ and the target point $p_{t}$. The angles $\theta$ and $\phi$ represent the deviations from the $y$-axis and $x$-axis, respectively. 
Without paraxial approximations, the Rayleigh-Sommerfeld diffraction integral calculates complex amplitude at any target point $(\rho, \theta, \phi)$ by integrating the effect of every source point $(x', y')$ located at distance $r$ from target point, where each single position on metasurface is assumed as Huygens’ source emitting spherical wave of wavelength $\lambda$ and wavenumber $k$. However, the direct numerical integration of every source point is computationally inefficient and hard to be adopted in a first-order gradient optimizer due to extravagant computing resources and time. 

Therefore, we reformulate equation~\eqref{eq:sph_coord} to enable its efficient evaluation with a Fourier transform as:
\begin{align}
\label{eq:prop_ours}
  U(\rho, \theta, \phi)  &=  \tau
 \int\int U'(x', y')e^{-2\pi j(\frac{\sin{\theta}\cos{\phi}}{\lambda}x' + \frac{\cos{\theta}}{\lambda}y')}\, dx'dy',
\end{align}
where $\tau$ is $\frac{e^{jk\rho}}{\lambda j}\frac{\sin{\theta}\sin{\phi}}{\rho}$, which models intensity drops at larger diffraction angle and greater distance. Here, we presume that the propagation distance is sufficiently larger than the wavelength of light, and the physical size of metasurface. 

Then, we can evaluate equation~\eqref{eq:prop_ours} with a FFT and reparameterization as
\begin{align}
    \label{eq:prop_f_ours}
    \mathbf{U}  = \boldsymbol{\tau} \odot f_\text{reparam}\left( \mathcal{F} \{\mathbf{U}'\} \right),
\end{align}
where $\mathbf{U}$ and $\boldsymbol{\tau}$ are the discrete variables of $U$ and $\tau$ with respect to the spherical coordinate $(\theta,\phi)$ for a specific value of $\rho$. For the source wavefront $U'$, we denote $\mathbf{U}'$ as its discrete variable, which is defined on the Cartesian coordinate $(x',y')$.
$\mathcal{F}$ is the discrete FFT operator and $f_\text{reparam}$ reparameterizes the transformed output to the spherical coordinate $(\theta,\phi)$.
$\odot$ is Hadamard product.

Our propagation model shows computational efficiency, outperforming full-space Rayleigh-Sommerfeld diffraction integral with a speed-up of over 50,000$\times$ at a resolution of 512$\times$512, and can accurately model 180$^\circ$ wavefront propagation without paraxial approximation.  
Fig.~\ref{fig:360illum}a illustrates the computational process of evaluating our propagation model.
See Supplementary Notes 5, 3, 4, and 6 for a detailed derivation, comparative discussion with other structured-light methods and wave propagators, and applicable range of the propagation model, respectively.

We achieve 360$^\circ$ structured light using our differentiable wave propagation model as well as the exploitation of the light transmitted and reflected by a metasurface, thus covering a full 360$^\circ$ space.
We design metasurface with rectangular-shaped meta-atoms (Fig.~\ref{fig:360illum}b). Given circularly-polarized light incident on the meta-atom, where the in-plane rotation angle of the meta-atom is $\theta_m$, both the transmitted and reflected wavefronts of cross-polarized light undergo phase delay of $\pm2\theta_m$, where + and - for the right circularly, and left circularly polarized light incident, respectively. 
Here, it should be noted that due to their same phase modulation, the structured-light pattern at transmission and reflection are identical. 
In terms of amplitude, they change by the modulation factors of $\frac{t_l-t_s}{2} $ and $\frac{r_l-r_s}{2}$, respectively, where $t_{l}$ and $t_{s}$ ($r_{l}$ and $r_{s}$) represent complex transmission (reflection) coefficients along the long and short axes of the meta-atom. By manipulating the geometry of the meta-atoms to induce large values of cross-polarization efficiencies $T_\text{cross} = \left| \frac{t_l-t_s}{2} \right|^2$ and $R_\text{cross} = \left| \frac{r_l-r_s}{2} \right|^2$ through rigorous coupled-wave analysis, we ensure achieving high-efficiency 360$^\circ$ structured light on transmission and reflection (Fig.~\ref{fig:360illum}c).

We design neural 360$^\circ$ structured light by learning the metasurface design using a first-order optimizer for downstream tasks using the differentiable framework.
Specifically, we optimize the 2D phase map of the metasurface, denoted as $\Phi$, utilized by a downstream reconstructor $f_\text{task}$ to compute the loss function $\mathcal{L}$, reflecting the end goal of the downstream task:
\begin{equation}
\label{eq:opt_meta}
\underset{\Phi}{\text{minimize}} \quad \mathcal{L}(f_\text{task}(f_\text{prop}(\Phi))),
\end{equation}
where $f_\text{prop}(\Phi)$ denotes the propagated 360$^\circ$ wavefront for a given source wavefront modulated by a metasurface with phase map $\Phi$, as described by equation~\eqref{eq:prop_f_ours}. The reconstruction loss is then backpropagated to the metasurface phase map $\Phi$ using a first-order gradient optimizer.

As a proof-of-concept of applying neural 360$^\circ$ structured light, we optimize the phase map of metasurfaces $\Phi$ to produce a high-fidelity 360$^\circ$ light projection of a desired holographic image $Y$ by solving the following optimization problem:
\begin{equation}
\underset{\Phi}{\text{minimize}} \quad \| |f_\text{prop}(\Phi)|^2-Y \|_2^2,
\end{equation}
where holographic image reconstructor $f_\text{task}$ computes mean square error (MSE) between target intensity distribution $Y$ and the simulated structured light. 
The metasurface with the learned phase map is fabricated (Fig.~\ref{fig:360illum}d) and used for realizing high-fidelity holographic 360$^\circ$ light projection (Fig.~\ref{fig:360illum}e, f). See Supplementary Figure 5 for additional experimental demonstration of 360$^\circ$ light projection.

\subsection{3D imaging with neural 360$^\circ$ structured light.}
Active stereo is a 3D imaging modality that employs multi-view cameras coupled with a structured-light module. The projected structured light serves as effective features for correspondence matching, enabling reliable multi-view 3D imaging even for scenes with complex visual texture and geometry\cite{hartley2003multiple}.
Here, we leverage neural 360$^\circ$ structured light for the first 360$^\circ$ active-stereo 3D imaging. We jointly optimize the metasurface phase map $\Phi$ and the parameters of a depth-reconstruction neural network $\Theta$, yielding an optimized 360$^\circ$ structured light suitable for 3D imaging.
The overall framework is illustrated in Fig.~\ref{fig:e2e}a.

Using our wave propagation method, fish-eye camera models, and multi-view geometry, we develop a differentiable image formation model that simulates an image captured by a fisheye camera under the neural 360$^\circ$ structured light. The image formation of each fisheye image is defined as
\begin{equation}
\label{eq:image_formation}
I = f_\text{clip}\left( S \odot R \odot \left( O \odot  f_\text{warp}( \alpha f_\text{prop}(\Phi), D ) + \beta \right) + \eta \right),
\end{equation}
where $f_\text{clip}$ bounds the pixel value with the maximum value of one, $S$ denotes the cosine foreshortening term, $R$ represents the scene reflectance, $O$ is the occlusion map indicating the visibility of the structured light at each camera pixel, $f_\text{warp}$ transforms the structured-light image from the viewpoint of the metasurface to the camera viewpoint via backward mapping from a camera-image pixel to an illumination-image pixel (Fig.~\ref{fig:e2e}b), and $D$ is the depth map. 
Parameters $\alpha$, $\beta$, and $\eta$ account for RGB-channel-dependent laser intensity illuminating the metasurface, intensity from ambient illumination, and Gaussian noise, respectively.
Utilizing the image formation of each fisheye image, our image formation model, $f_\text{render}$, simulates four fisheye images:
\begin{equation}
\label{eq:image_formation_four}
\{I_\text{left-front},I_\text{right-front},I_\text{left-rear},I_\text{right-rear}\} = f_\text{render}\left( \Phi, S, R, O, D\right),
\end{equation}
where $I_\text{left-front}$ and $I_\text{right-front}$ correspond to the left and right images for the frontal view, while $I_\text{left-rear}$ and $I_\text{right-rear}$ represent the rear-view images.
Further details on image formation can be found in Supplementary Note 7.

Subsequently, we employ these four fisheye images to reconstruct the 360$^\circ$ depth map. 
For the frontal view, we extract features from the left and right fisheye images using a convolutional encoder $f_\text{feat}$.
Then, we construct a cost volume with spherical sweeping $f_\text{sphere}$ on the extracted features and infer the depth map with an edge-aware refinement $f_\text{edge}$. 
This process is expressed as
\begin{equation}
 \hat{D}_\text{front} = f_\text{edge}\left(f_\text{sphere}\left(f_\text{feat} (I_\text{right-front} ), f_\text{feat}(I_\text{left-front}) \right), I_\text{right-front}\right).
 \end{equation}
 By applying a similar process to the rear view using the left and right images $I_\text{left-rear}, I_\text{right-rear}$, we can obtain the rear-view depth map $\hat{D}_\text{rear}$, thereby forming a 360$^\circ$ depth map $\hat{D}$.
In summary, our depth reconstructor, denoted as $f_\text{depth}$, estimates the 360$^\circ$ depth map $\hat{D}$ using the four fisheye images:
 \begin{equation}
\label{eq:image_formation_four}
\hat{D} = f_\text{depth}\left(I_\text{left-front},I_\text{right-front},I_\text{left-rear},I_\text{right-rear}\right).
\end{equation}
More details on the network architecture can be found in  Supplementary Note 8.

Leveraging our differentiable framework of the image formation and the depth reconstruction, we learn the metasurface phase map $\Phi$ and the parameters of the depth reconstructor $\Theta$ by minimizing the loss function $\mathcal{L}$ over a synthetic training dataset composed of reflectance, normal, occlusion, and depth maps:
\begin{equation}
\label{eq:opt_3d}
 \underset{\Phi, \Theta}{\text{minimize}} \,  \sum_{i=1}^N \mathcal{L}\left( f_\text{depth}(f_\text{render}(\Phi, S_i, R_i, O_i, D_i); \Theta), D_i \right),
\end{equation}
where $S_i, R_i, O_i, D_i$ are the foreshortening, reflectance, occlusion, and depth maps for $i$-th training sample. $N$ is the number of training samples.

To this end, we present the 360$^\circ$ synthetic dataset of reflectance, normal, occlusion, and depth maps rendered by a computer-graphics path tracer\cite{blender}.
Using the dataset, we solve equation~\eqref{eq:opt_3d} with a two-stage training strategy, promoting high contrast of structured-light patterns in the early training iterations to enhance far-depth imaging and later optimizing the depth-estimation neural network only. 
The optimization process over training iterations is shown in Fig.~\ref{fig:e2e}c and Supplementary Video.
Further details on the training details can be found in Supplementary Note 9.



We have identified two distinct attributes from the learned neural 360$^\circ$ structured light, designed for 3D imaging, as depicted in Fig.~\ref{fig:e2e}d.
In contrast to multi-dot patterns\cite{kim2022metasurface} characterized by evenly-spaced dots of consistent intensity, our learned pattern manifests a range of non-uniform intensities.
Secondly, our pattern offers intricate features that do not align directly with pixel locations for depth candidates (Fig.~\ref{fig:e2e}e).
This distinctiveness could provide unique matching cues for depth estimation, as compared to the regularly-positioned dots of the heuristic multi-dot patterns\cite{kim2022metasurface}.

\subsection{Experimental demonstration of 3D imaging.}
We fabricate a metasurface with the learned phase map $\Phi$ to demonstrate 360$^\circ$ 3D imaging. Specifically, as illustrated in Fig.~\ref{fig:demonstration}a, we construct an experimental prototype integrating a metasurface illumination module and four fisheye cameras.
Our approach to 360$^\circ$ structured light leverages both transmission and reflection channels of the metasurface through the integration of polarization optical elements. A half-wave plate (HWP) alters the axis of linearly-polarized laser light to maximizes the reflected energy from a polarizing beam splitter (PBS). The now horizontally-linearly polarized light undergoes a state change to left circular polarization (LCP) upon passing through a quarter-wave plate (QWP), incident on the metasurface.
In the reflection channel of the metasurface, the LCP converts to right circular polarization (RCP). Uniquely, some of the reflected light near the optical axis re-traverses the QWP and PBS, but in the opposite direction. The polarization state undergoes transformation upon traversing the QWP, ultimately transmitting through the PBS.
Given the angle of incidence range and the physical size of the QWP and PBS, our prototype is capable of illuminating the reflective regime without distortion. See Supplementary Note 1 for detailed experimental setup. This results in a full 360$^\circ$ structured light, while maintaining the level of detail mirrored in the simulations (Fig.~\ref{fig:demonstration}b).
We capture four fish-eye images ${I_\text{left-front}, I_\text{right-front}, I_\text{left-rear}, I_\text{right-front}}$ using our learned metasurface-enabled 360$^\circ$ structured light (Fig.~\ref{fig:demonstration}c). Subsequently, we apply our depth reconstruction, to obtain the front and rear depth maps.
To experimentally assess 3D imaging with neural 360$^\circ$ structured light, we measured the depth of texture-less planes situated at six different depths across the frontal FoV range. As demonstrated in Fig.~\ref{fig:demonstration}d, our imaging system delivers accurate depth estimation consistently across different depths with the average depth error of 35\,mm over 2.5\,m ranges, outperforming conventional 360$^\circ$ structured light\cite{kim2022metasurface} by 5.09$\times$ lower depth-reconstruction RMSE. 
Fig.~\ref{fig:demonstration}e shows qualitative results of two real-world scenes.
Depth of real-world objects, including floors, human, furniture, and statues, are accurately reconstructed. 
In synthetic test scenes, our neural 360$^\circ$ structured light outperforms using the heuristically-designed 360$^\circ$ structured light including the super-cell method\cite{kim2022metasurface} and the pattern generated by randomized phase distribution of metasurface in terms of depth accuracy by 1.86$\times$ in RMSE.
Additional results on experimental captures and synthetic scenes are provided in Supplementary Notes 11 and 12.



\section*{Discussion}
In this work, we introduce neural 360$^\circ$ structured light, facilitated by metasurface-aided illumination. This metasurface is directly optimized for a downstream task through a differentiable framework that includes full-space wave propagation, task-specific image formation, and reconstruction. We provide both simulated and experimental demonstrations of the application of neural 360$^\circ$ structured light for 3D imaging and holographic light projection. Regarding 3D imaging, we obtained accurate depth reconstruction using neural 360$^\circ$ structured light, achieving a depth-reconstruction root mean squared error 5.09$\times$ lower than that obtained from using heuristically-designed 360$^\circ$ structured light\cite{kim2022metasurface}. For holographic light projection, we succeeded in creating a high-fidelity 360$^\circ$ light projection of complex patterns, an accomplishment not yet reported elsewhere.
While these findings are promising, there are several aspects of neural 360$^\circ$ structured light that can be further improved. First, our current single-layer metasurface design replicates transmission and reflection structured-light patterns. This limitation could be overcome with two metasurfaces each designed for transmission and reflection, which could enable a 360$^\circ$ holographic projection without the replication constraint. Second, the incorporation of dynamic phase-modulation elements such as SLMs along with metasurfaces could allow dynamically-adjustable 360$^\circ$ structured light, which would be advantageous for 360$^\circ$ video holographic display and environment-adaptive 360$^\circ$ 3D imaging. Third, our current experimental prototype could be enhanced through the use of beam splitter-based fisheye cameras and by operating at near-infrared wavelengths, thereby providing compact imaging that is non-invasive to human vision. 
Lastly, our wave propagation model could be extended to incorporate near-field full-space light propagation, potentially enabling 3D 360$^\circ$ holographic display.
We believe that neural 360$^\circ$ structured light is a significant step towards compact and robust 360$^\circ$ imaging and display, holding potential benefits for a range of applications, including robotics, augmented-reality systems, human-computer interaction, and autonomous vehicles.


\begin{methods}

\subsection{Optimization.}
We used PyTorch to develop optimization for neural 360$^\circ$ structured light. 
A single NVIDIA A6000 GPU with 48GB memory is used for training and evaluation. See Supplementary Note 9 for training strategy, hyperparameters, and loss functions.

\subsection{Dataset.}
For the training of the metasurface and the reconstructor in neural 360$^\circ$ 3D imaging, we construct a 360$^\circ$ dataset consisting of reflectance, depth, normal, and occlusion masks for 200 synthetic scenes. These scenes are generated using Blender\cite{blender} and rendered with four virtual fish-eye cameras with intrinsic and extrinsic parameters mirroring those of our experimental prototype. For the scene construction, we utilize 30 distinct objects, each with unique shapes and textures randomly sampled from the ShapeNet\cite{shapenet2015}. These objects are placed at varying depths within a range of 0.3\,m to 3\,m.
For background, we use cube and spherical objects positioned at depths ranging from 3\,m to 5\,m.

\subsection{Experimental Setup.}
We construct an imaging system shown in Fig.~\ref{fig:demonstration}a, consisting of illumination and capture modules. In the illumination module, we utilize a linearly-polarized laser at a wavelength of 532\,nm (MGL-DS-532). We use a HWP (Thorlabs WPH10M-532), PBS (Edmund \#48-545), and QWP (Thorlabs WPQ10E-532) to alter polarization states. For the capture module, we employ four RGB cameras (two of Basler a2A1920-160ucBAS and two of Basler a2A1920-160ucPRO) equipped with four fish-eye lenses, resulting in 185$^\circ$ FoV. The entire system was mounted on a $300 \times 150$\,mm$^2$ breadboard, resulting in a compact configuration.

\subsection{Metasurface Fabrication.}
On a glass substrate, 246\,nm thick hydrogenated amorphous silicon (a-Si:H) was deposited by plasma enhanced chemical vapor deposition (PECVD, BMR Technology HiDep-SC). Then, metasurface pattern was transferred to spin-coated polymethyl methacrylate (PMMA, 495 A6) photoresist using electron beam lithography (EBL, ELIONIX ELS-7800). Exposed pattern was developed using MIBK/IPA 1:3 developer, and 35 nm thick chromium (Cr) was deposited using electron beam evaporator (EBE, KVT KVE-ENS4004). After lift-off process, the metasurface pattern was finally transferred to the a-Si:H through dry-etching process (DMS, silicon/metal hybrid etcher) using Cr as etching mask. Remaining Cr mask was removed using Cr etchant (CR-7).

\subsection{Data Availability.}
Our 360$^\circ$ synthetic dataset and the learned metasurface phase map will be made publicly available on GitHub.

\subsection{Code Availability.}
The code used to generate the findings of this study will be made public on GitHub.

\end{methods}


\bibliographystyle{naturemag}
\bibliography{reference}


\begin{addendum}
	\item 
    S.-H.B. acknowledges the National Research Foundation of Korea (NRF) grant (NRF-2022R1A6A1A03052954) funded by the Ministry of Education (2022R1A6A1A03052954), the Samsung Research Funding \& Incubation Center for Future Technology grant (SRFC-IT1801-52) funded by Samsung Electronics, and the Alchemist Project NTIS1415187366 (20025752) funded By the Ministry of Trade, Industry \& Energy (MOTIE, Korea).
    J.R. acknowledges the Samsung Research Funding \& Incubation Center for Future Technology grant (SRFC-IT1901-52) funded by Samsung Electronics, and the National Research Foundation (NRF) grant (NRF-2022M3C1A3081312) funded by the Ministry of Science and ICT (MSIT) of the Korean government. G.K. acknowledges the POSTECH Alchemist fellowship.

	\item [Author Contributions]
 S.-H.B. and E.C. conceived the idea. E.C. designed the propagation model. E.C., G.K., and J.Y. verified the propagation model. E.C. and Y.J. performed end-to-end training and synthetic experiments. E.C., G.K., and Y.J. implemented the experimental prototype. G.K. fabricated the devices. J.R. guided the material characterization and device fabrication. All authors participated in discussions and contributed to writing the manuscript. S.-H.B. and J.R. guided all aspects of the work.
	\item [Competing Interests] The authors declare no competing financial interests.
	\item [Supplementary Information] Supplementary Information accompanies this manuscript as part of the submission files.
	\item [Correspondence] Correspondence should be addressed to S.-H.B. or J.R.
\end{addendum}

\clearpage

\begin{figure}[t]
	\centering
		\includegraphics[width=\columnwidth]{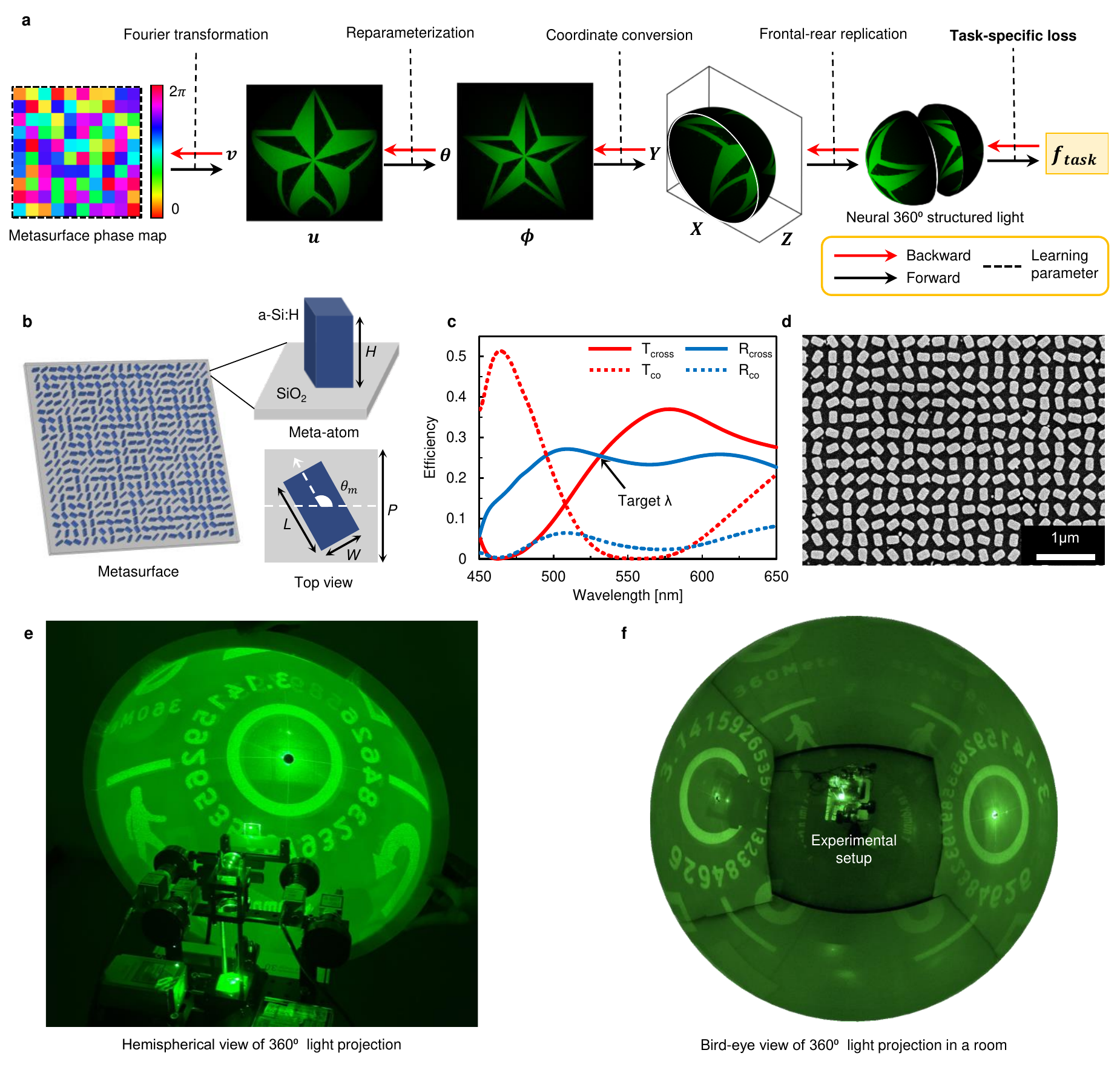}
\caption{Neural 360$^\circ$ structured light with learned metasurface \textbf{a} Design of neural 360$^\circ$ structured light. The phase map of the metasurface is optimized for a specific downstream task with our wave propagation model. \textbf{b} Illustration of the geometric phase-based metasurface composed of hydrogenated amorphous silicon (a-Si:H) rectangular meta-atoms on a silica substrate. The unit cell design of the meta-atom is represented by a set of geometric parameters, height (H), length (L), width (W), and pitch (P). \textbf{c} Spin-converted and spin-preserved efficiencies of the optimized meta-atom with H = 246 nm, L = 180 nm, W = 100 nm, and P = 260 nm. The geometric parameters of the meta-atom are optimized to have high conversion efficiencies in both transmission and reflection regime at a target wavelength, thereby ensuring 360$^\circ$ projection. \textbf{d} Scanning electron microscopy (SEM) image of the fabricated metasurface for holographic 360$^\circ$ light projection. \textbf{e, f} Experimental demonstration of 360$^\circ$ light projection with neural 360$^\circ$ structured light in a hemispherical screen and a confined room.
}

		\label{fig:360illum}
\end{figure}

\begin{figure}[t]
	\centering
		\includegraphics[width=\columnwidth]{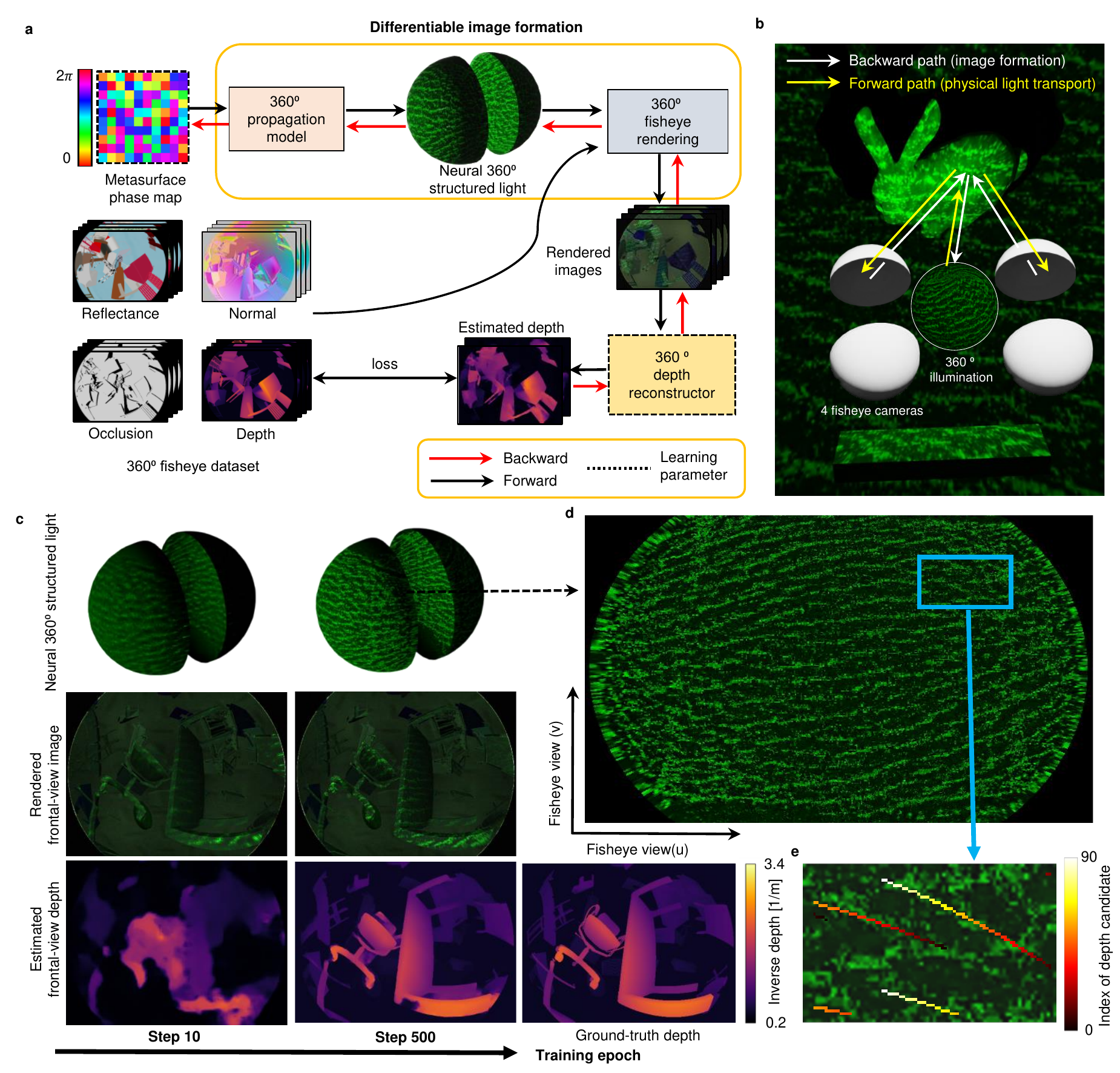}
\caption{3D imaging with neural 360$^\circ$ structured light. \textbf{a} Schematic of the differentiable framework of 360$^\circ$ wave propagation model, fisheye rendering, and depth reconstructor. The metasurface phase map and the neural network parameters are jointly learned using our synthetic 360$^\circ$ dataset that includes reflectance, normal, occlusion, and depth maps. \textbf{b} In our fisheye rendering, we use backward mapping from a camera pixel to a corresponding pixel in the 360$^\circ$ structured-light image. \textbf{c} Visualization of the change of neural 360$^\circ$ structured light, simulated fisheye image, and estimated depth map over training iterations. In the early iterations, the overall shape of the structured-light pattern appears, and the density and pattern intensity with high-frequency details emerge as training proceeds. \textbf{d} The learned neural 360$^\circ$ structured light presents non-uniform intensity and slanted patterns. \textbf{e} These unique characteristics serve as effective cues for depth estimation by furnishing distinctive correspondence-matching features. 
 }
		\label{fig:e2e}
\end{figure}

\begin{figure}[t]
	\centering
		\includegraphics[width=\columnwidth]{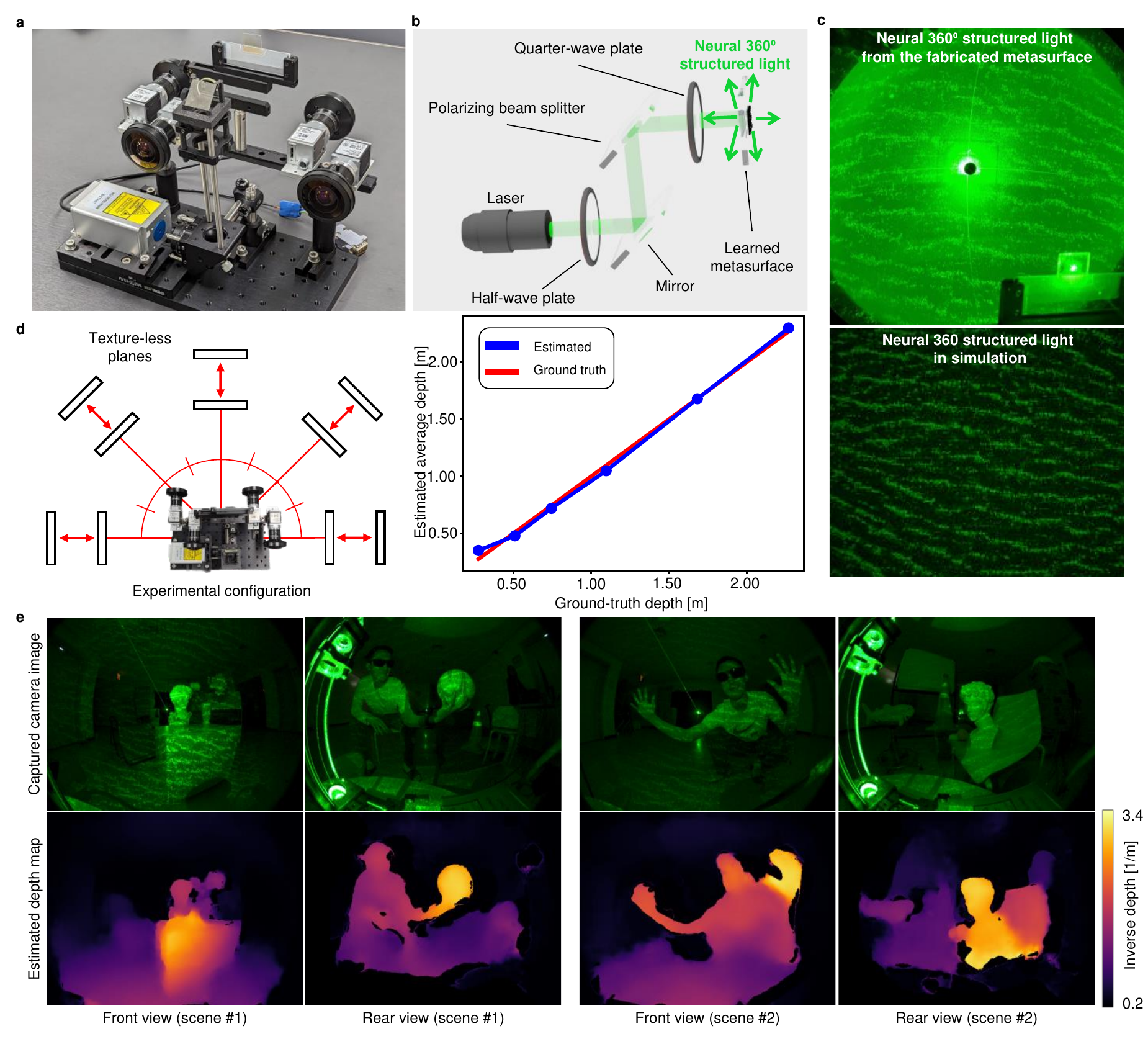}
\caption{Experimental demonstration of 360$^\circ$ 3D imaging \textbf{a} We experimentally demonstrate 3D imaging with neural 360$^\circ$ structured light by constructing a prototype composed of an illumination module and four fisheye cameras. \textbf{b} For the illumination module, we use the fabricated metasurface with the phase map learned from training. Laser illumination passes through multiple polarizing optics before impinging on the metasurface. Transmitted and reflected light from the metasurface generates the learned 360$^\circ$ structured light. \textbf{c} We show the learned metasurface illumination in simulation and from the fabricated metasurface. \textbf{d} We assess the 3D imaging capability of the experimental system by capturing texture-less planes at multiple distances within the hemispherical space, resulting in accurate depth reconstruction with an average error of 3.5 cm. \textbf{e} We show experimental captures of two different scenes containing various real-world objects, including a human subject. The captured images and estimated frontal and rear depth maps demonstrate the effectiveness of the proposed system.
}
		\label{fig:demonstration}
\end{figure}

\end{document}


\maketitle

\begin{affiliations}
  \item Department of Computer Science and Engineering, Pohang University of Science and Technology (POSTECH), Pohang 37673, Republic of Korea
 \item Department of Mechanical Engineering, Pohang University of Science and Technology (POSTECH), Pohang 37673, Republic of Korea
 \item Department of Chemical Engineering, Pohang University of Science and Technology (POSTECH), Pohang 37673, Republic of Korea
 \item POSCO-POSTECH-RIST Convergence Research Center for Flat Optics and Metaphotonics, Pohang 37673, Republic of Korea
 \item National Institute of Nanomaterials Technology (NINT), Pohang 37673, Republic of Korea
   \item [$*$] Equal contribution
 \item [$\dagger$] Corresponding author. E-mail: shwbaek@postech.ac.kr, jsrho@postech.ac.kr
\end{affiliations}


\definecolor{brightray}{rgb}{0.8,0.8,0.8}
\definecolor{Gray}{rgb}{0.5,0.5,0.5}
\definecolor{darkblue}{rgb}{0,0,0.7}
\definecolor{orange}{rgb}{1,.5,0} 
\definecolor{red}{rgb}{1,0,0} 
\definecolor{blue}{rgb}{0,0,1} 
\definecolor{darkgreen}{rgb}{0,0.7,0} 
\definecolor{darkred}{rgb}{0.7,0,0} 

\newcommand{\heading}[1]{\noindent\textbf{#1}}
\newcommand{\note}[1]{{{\textcolor{orange}{#1}}}}
\newcommand{\todo}[1]{{\textcolor{red}{TODO: #1}}}
\newcommand{\changed}[1]{{\textcolor{blue}{#1}}}
\newcommand{\removed}[1]{{\textcolor{brightray}{{#1}}}}
\newcommand{\revision}[1]{{{#1}}}
\newcommand{\place}[1]{ \begin{itemize}\item\textcolor{darkblue}{#1}\end{itemize}}
\newcommand{\de}{\mathrm{d}}

\newcommand{\BEAS}{\begin{eqnarray*}}
\newcommand{\EEAS}{\end{eqnarray*}}
\newcommand{\BEA}{\begin{eqnarray}}
\newcommand{\EEA}{\end{eqnarray}}
\newcommand{\BEQ}{\begin{equation}}
\newcommand{\EEQ}{\end{equation}}
\newcommand{\BIT}{\begin{itemize}}
\newcommand{\EIT}{\end{itemize}}
\newcommand{\BNUM}{\begin{enumerate}}
\newcommand{\ENUM}{\end{enumerate}}

\newcommand{\BA}{\begin{array}}
\newcommand{\EA}{\end{array}}

\newcommand{\eg}{{\it e.g.}}
\newcommand{\ie}{{\it i.e.}}
\newcommand{\etc}{{\it etc.}}

\newcommand{\ones}{\mathbf 1}

\newcommand{\reals}{{\mbox{\bf R}}}
\newcommand{\integers}{{\mbox{\bf Z}}}
\newcommand{\eqbydef}{\mathrel{\stackrel{\Delta}{=}}}
\newcommand{\complex}{{\mbox{\bf C}}}
\newcommand{\symm}{{\mbox{\bf S}}}  

\newcommand{\Span}{\mbox{\textrm{span}}}
\newcommand{\Range}{\mbox{\textrm{range}}}
\newcommand{\nullspace}{{\mathcal N}}
\newcommand{\range}{{\mathcal R}}
\newcommand{\Nullspace}{\mbox{\textrm{nullspace}}}
\newcommand{\Rank}{\mathop{\bf Rank}}
\newcommand{\Tr}{\mathop{\bf Tr}}
\newcommand{\diag}{\mathop{\bf diag}}
\newcommand{\lambdamax}{{\lambda_{\rm max}}}
\newcommand{\lambdamin}{\lambda_{\rm min}}

\newcommand{\Expect}{\mathop{\bf E{}}}
\newcommand{\Prob}{\mathop{\bf Prob}}
\newcommand{\erf}{\mathop{\bf erf}}

\newcommand{\Co}{{\mathop {\bf Co}}}
\newcommand{\co}{{\mathop {\bf Co}}}
\newcommand{\dist}{\mathop{\bf dist{}}}
\newcommand{\Ltwo}{{\bf L}_2}
\newcommand{\QED}{~~\rule[-1pt]{8pt}{8pt}}\def\qed{\QED}
\newcommand{\approxleq}{\mathrel{\smash{\makebox[0pt][l]{\raisebox{-3.4pt}{\small$\sim$}}}{\raisebox{1.1pt}{$<$}}}}
\newcommand{\epi}{\mathop{\bf epi}}

\newcommand{\vol}{\mathop{\bf vol}}
\newcommand{\Vol}{\mathop{\bf vol}}
\newcommand{\Card}{\mathop{\bf card}}

\newcommand{\dom}{\mathop{\bf dom}}
\newcommand{\aff}{\mathop{\bf aff}}
\newcommand{\cl}{\mathop{\bf cl}}
\newcommand{\Angle}{\mathop{\bf angle}}
\newcommand{\intr}{\mathop{\bf int}}
\newcommand{\relint}{\mathop{\bf rel int}}
\newcommand{\bd}{\mathop{\bf bd}}
\newcommand{\vect}{\mathop{\bf vec}}
\newcommand{\dsp}{\displaystyle}
\newcommand{\foequal}{\simeq}
\newcommand{\VOL}{{\mbox{\bf vol}}}
\newcommand{\xopt}{x^{\rm opt}}

\newcommand{\Xb}{{\mbox{\bf X}}}
\newcommand{\xst}{x^\star}
\newcommand{\varphist}{\varphi^\star}
\newcommand{\lambdast}{\lambda^\star}
\newcommand{\Zst}{Z^\star}
\newcommand{\fstar}{f^\star}
\newcommand{\xstar}{x^\star}
\newcommand{\xc}{x^\star}
\newcommand{\lambdac}{\lambda^\star}
\newcommand{\lambdaopt}{\lambda^{\rm opt}}

\newcommand{\geqK}{\mathrel{\succeq_K}}
\newcommand{\gK}{\mathrel{\succ_K}}
\newcommand{\leqK}{\mathrel{\preceq_K}}
\newcommand{\lK}{\mathrel{\prec_K}}
\newcommand{\geqKst}{\mathrel{\succeq_{K^*}}}
\newcommand{\gKst}{\mathrel{\succ_{K^*}}}
\newcommand{\leqKst}{\mathrel{\preceq_{K^*}}}
\newcommand{\lKst}{\mathrel{\prec_{K^*}}}
\newcommand{\geqL}{\mathrel{\succeq_L}}
\newcommand{\gL}{\mathrel{\succ_L}}
\newcommand{\leqL}{\mathrel{\preceq_L}}
\newcommand{\lL}{\mathrel{\prec_L}}
\newcommand{\geqLst}{\mathrel{\succeq_{L^*}}}
\newcommand{\gLst}{\mathrel{\succ_{L^*}}}
\newcommand{\leqLst}{\mathrel{\preceq_{L^*}}}
\newcommand{\lLst}{\mathrel{\prec_{L^*}}}

\newtheorem{theorem}{Theorem}[section]
\newtheorem{corollary}{Corollary}[theorem]
\newtheorem{lemma}[theorem]{Lemma}
\newtheorem{proposition}[theorem]{Proposition}

\newenvironment{algdesc}%
{\begin{quote}}{\end{quote}}

\def\figbox#1{\framebox[\hsize]{\hfil\parbox{0.9\hsize}{#1}}}

\makeatletter
\long\def\@makecaption#1#2{
   \vskip 9pt
   \begin{small}
   \setbox\@tempboxa\hbox{{\bf #1:} #2}
   \ifdim \wd\@tempboxa > 5.5in
        \begin{center}
        \begin{minipage}[t]{5.5in}
        \addtolength{\baselineskip}{-0.95pt}
        {\bf #1:} #2 \par
        \addtolength{\baselineskip}{0.95pt}
        \end{minipage}
        \end{center}
   \else
    \hbox to\hsize{\hfil\box\@tempboxa\hfil}
   \fi
   \end{small}\par
}
\makeatother

\newcounter{oursection}
\newcommand{\oursection}[1]{
 \addtocounter{oursection}{1}
 \setcounter{equation}{0}
 \clearpage \begin{center} {\Huge\bfseries #1} \end{center}
 {\vspace*{0.15cm} \hrule height.3mm} \bigskip
 \addcontentsline{toc}{section}{#1}
}
\newcommand{\oursectionf}[1]{  
 \addtocounter{oursection}{1}
 \setcounter{equation}{0}
 \foilhead[-.5cm]{#1 \vspace*{0.8cm} \hrule height.3mm }
 \LogoOn
}
\newcommand{\oursectionfl}[1]{  
 \addtocounter{oursection}{1}
 \setcounter{equation}{0}
 \foilhead[-1.0cm]{#1}
 \LogoOn
}

\newcommand{\Mat}[1]    {{\ensuremath{\mathbf{\uppercase{#1}}}}} 
\newcommand{\Vect}[1]   {{\ensuremath{\mathbf{\lowercase{#1}}}}} 
\newcommand{\Vari}[1]   {{\ensuremath{\mathbf{\lowercase{#1}}}}} 
\newcommand{\Id}				{\mathbb{I}} 
\newcommand{\Diag}[1] 	{\operatorname{diag}\left({ #1 }\right)} 
\newcommand{\Opt}[1] 	  {{#1}_{\text{opt}}} 
\newcommand{\CC}[1]			{{#1}^{*}} 
\newcommand{\Op}[1]     {\Mat{#1}} 
\newcommand{\mini}[1] {{\mbox{argmin}}_{#1} \: \: } 
\newcommand{\argmin}[1] {\underset{{#1}}{\mathop{\rm argmin}} \: \: } 
\newcommand{\argmax}[1] {\underset{{#1}}{\mathop{\rm argmax}} \: \: } 
\newcommand{\minimize}{\mathop{\rm minimize} \: \:}
\newcommand{\minimizeu}[1]{\underset{{#1}}{\mathop{\rm minimize}} \: }
\newcommand{\grad}      {\nabla}
\newcommand{\kron}{\otimes} 

\newcommand{\gradt}     {\grad_\z}
\newcommand{\gradx}     {\grad_\x}
\newcommand{\Drv}     	{\Mat{D}} 
\newcommand{\step}      {\text{\textbf{step}}}
\newcommand{\prox}[1]   {\mathbf{prox}_{#1}}
\newcommand{\ind}[1]    {\operatorname{ind}_{#1}}
\newcommand{\proj}[1]   {\Pi_{#1}}
\newcommand{\pointmult}{\odot} 
\newcommand{\rr}   {\mathcal{R}}

\newcommand{\Basis}{\Mat{D}}         		
\newcommand{\Corr}{\Mat{C}}             
\newcommand{\conv}{\ast} 
\newcommand{\meas}{\Vect{b}}            
\newcommand{\Img}{I}                    
\newcommand{\img}{\Vect{i}}             
\newcommand{\vv}{\Vect{v}}
\newcommand{\p}{\Vect{p}}
\newcommand{\Splitvar}{T}                
\newcommand{\splitvar}{\Vect{t}}         
\newcommand{\Splitbasis}{J}                
\newcommand{\splitbasis}{\Vect{j}}         
\newcommand{\var}{\Vari{z}}

\newcommand{\FT}[1]			{\mathcal{F}\left( {#1} \right)} 
\newcommand{\IFT}[1]			{\mathcal{F}^{-1}\left( {#1} \right)} 

\newcommand{\func}{f}
\newcommand{\fMat}{\Mat{K}}

\newcommand{\avar}{\Vari{v}}
\newcommand{\aspvar}{\Vari{z}}

\newcommand{\mask}{\Mat{M}}

\newcommand{\Pen}      		{F} 
\newcommand{\cardset}     {\mathcal{C}}
\newcommand{\Dat}      		{G} 
\newcommand{\Reg}      		{\Gamma} 

\newcommand{\Trans}{\mathbf{\uppercase{T}}} 
\newcommand{\Ph}{\mathbf{\uppercase{\Phi}}} 

\newcommand{\Tvec}{\Vect{T}} 
\newcommand{\Bvec}{\Vect{B}} 

\newcommand{\Wt}{\Mat{W}} 

\newcommand{\Perm}{\Mat{P}} 
\newcommand{\Cblur}{\Mat{C}} 

\newcommand{\DiagFactor}[1]     {\Mat{O}_{ #1 }}  

\newcommand{\Proj}{\Mat{P}}             

\newcommand{\Vector}[1]{\mathbf{#1}}
\newcommand{\Matrix}[1]{\mathbf{#1}}
\newcommand{\Tensor}[1]{\boldsymbol{\mathscr{#1}}}
\newcommand{\TensorUF}[2]{\Matrix{#1}_{(#2)}}

\newcommand{\MatrixKP}[1]{\Matrix{#1}_{\otimes}}
\newcommand{\MatrixKPN}[2]{\Matrix{#1}_{\otimes}^{#2}}

\newcommand{\MatrixKRP}[1]{\Matrix{#1}_{\odot}}
\newcommand{\MatrixKRPN}[2]{\Matrix{#1}_{\odot}^{#2}}

\newcommand{\HP}{\circ}
\newcommand{\HD}{\oslash}

\newcommand{\leftDB}{\left[ \! \left[}
\newcommand{\rightDB}{\right] \! \right]}

\newcommand{\transpose}{T}

\newcommand*\sstrut[1]{\vrule width0pt height0pt depth#1\relax}

\newcommand{\inlineeqnum}{\refstepcounter{equation}~~\mbox{(\theequation)}}
\newcommand{\eqname}[1]{\tag*{#1~(\theequation)}\refstepcounter{equation}}

\newcommand{\lambdas}{\boldsymbol{\lambda}}
\newcommand{\alb}{\boldsymbol{\alpha}} 	
\newcommand{\depth}{\boldsymbol{z}} 	
\newcommand{\albi}{\alpha} 	
\newcommand{\depthi}{z} 	
\newcommand{\ambient}{s}
\newcommand{\jitter}{w}
\newcommand{\z}{\Vect{z}} 							
\newcommand{\x}{\Vect{x}}             	
\newcommand{\y}{\Vect{y}}             	
\newcommand{\Kvar}{\Mat{K}}
\newcommand{\lagrangemult}{\boldsymbol{\nu}}
\newcommand{\scaledlagrange}{\Vect{u}}
\newcommand{\eps}{\epsilon}
\newcommand{\vp}{\Vect{v}}

\noindent In this supplementary document, we provide further discussion and results to support the main content of the manuscript.

\section*{Supplementary Video 1: End-to-end Optimization}
\label{sec:video_1}
Supplementary Video 1 shows our end-to-end learning process. The video consists of two parts.
First, it demonstrates the learned metasurface phase map and the corresponding structured-light pattern. As discussed in the main manuscript, the initial iterations of training establish the overall structure of the pattern, while subsequent iterations refine the finer details. This progression is visualized in the video.
Next, the video showcases the simulation of a captured image, the estimated depth map, and the ground truth during the training iterations. In the first stage of the training, the captured image evolves as the metasurface phase map is learned in conjunction with the depth estimation network. The video shows this dynamic process, highlighting the learned structured light.
In the second stage of training, the metasurface phase map is fixed, and the focus shifts solely to optimizing the depth estimation network. Consequently, the captured image remains consistent throughout this stage, while the quality of the estimated depth map progressively improves with the refined depth estimation network.

\clearpage

\section*{Supplementary Note 1: Experimental Setup}
\label{sec:setup}
\subsection{Polarization-aware illumination module.}
\begin{figure}[t]
	\centering
		\includegraphics[width=\columnwidth]{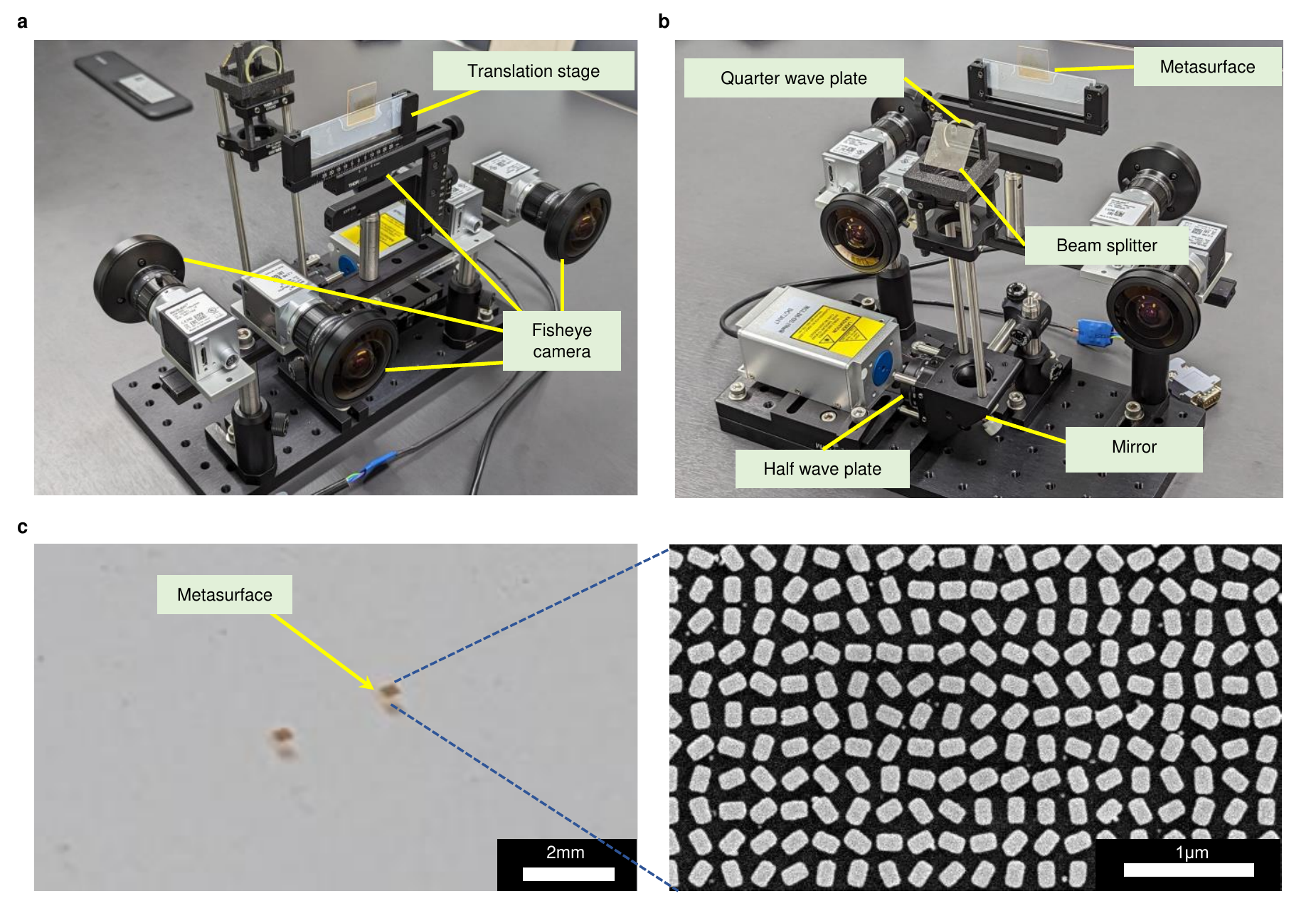}
		\caption{\textbf{a} Front view of the constructed prototype. \textbf{b} Rear view of the constructed prototype. Front and rear fisheye cameras capture transmitted and reflected neural structured light, respectively. \textbf{c} Photograph of the fabricted metasurface with a size of 260um x 260um (left) and the SEM image of the fabricated metasurface composed of meta-atoms with a length of 180 nm, width of 100 nm, and height of 246 nm (right).}
		\label{fig:setup}
\end{figure}

To experimentally demonstrate 360$^\circ$ 3D imaging based on end-to-end learned metasurface, we built a 360$^\circ$ illumination and image capturing prototype composed of metasurface and four fisheye cameras, two for front and the others for rear view (Fig.~\ref{fig:setup}a). To use both transmission and reflection channel of the metasurface, the illumination module consists of multiple polarization optical instruments including half-wave plate (HWP), polarizing beam splitter (PBS) and a quarter-wave plate (QWP) to minimize the area occluded by another components of prototype (Fig.~\ref{fig:setup}b). Following the Jones matrix representation, if we express the incoming light that has passed through the HWP and been reflected from the PBS as $\begin{bmatrix}
    1 \\
    1 \\
\end{bmatrix}$, then the polarization state of the input channel of metasurface is described as 
\begin{equation}
\begin{bmatrix}
    1   &    0 \\
    0   &   -i  \\
\end{bmatrix} \begin{bmatrix}
    1\\
    1\\
\end{bmatrix} = 
\begin{bmatrix}
    1 \\ -i\\
\end{bmatrix} : LCP
\end{equation}
which is then converted to RCP, $\begin{bmatrix}
    1 \\ i\\
\end{bmatrix}$, via learned metasurface and illuminates both the transmissive and reflective regimes. Here, 
$\begin{bmatrix}
    1   &  0 \\
    0   &  -i \\ 
\end{bmatrix}$ represents transformation matrix of the QWP. Unlike transmission channel of metasurface, the portion of reflected light near optical axis passes through the QWP and PBS again, but in the opposite direction. The polarization state after passing through the QWP is described as 
\begin{equation}
\begin{bmatrix}
    1   &    0 \\
    0   &   -i  \\
\end{bmatrix}^{-1} \begin{bmatrix}
    1\\
    i\\
\end{bmatrix} = 
\begin{bmatrix}
    1 \\-1\\
\end{bmatrix}
\end{equation}
which can transmit the PBS. Taking the range of angle of incidence and physical size of QWP and PBS in consideration, our prototype illuminates reflective regime without distortion or another inter-reflection, realizing full 360$^\circ$ illumination.

\subsection{Minimum distance between the metasurface and the quarter wave plate}
Due to the requirement of a stable performance for the polarization optical elements within a range of angles of incidence (AOI), there exists a lower limit to the displacement of the metasurface from the QWP and PBS. If the distance is reduced below this lower limit, some of the reflected light beams from the metasurface may exceed the permissible AOI, leading to undesired effect by the polarization optical elements. In our prototype design, to achieve maximum compactness, we consider an AOI range of used polarization elements, $\pm20^\circ$ for the QWP and $45\pm15^\circ$ for the PBS. Furthermore, we take into account the physical dimensions of the QWP ($2.54$\,cm) and the PBS ($2.5$\,cm). Consequently, to ensure the suppression of unwanted polarization-dependent distortion, the distances between the metasurface and the QWP, as well as the PBS, are chosen to exceed the calculated lower limit of approximately $4$\,cm.

\subsection{Alignment procedure}
Achieving accurate light projection with high-fidelity necessitates the maximization of laser energy conversion into the desired LCP. The objective can be accomplished by satisfying two conditions: 1) maximizing the reflected energy from the PBS, and 2) maximizing the ratio of energy converted to fully polarized light. To satisfy the first condition, the optical axis of the HWP is rotated to generate fully $45^\circ$ linearly polarized light. The second condition is met by rotating the optical axis of the QWP, resulting in the generation of fully LCP light. This adjustment effectively reduces the presence of partially polarized light, which otherwise degrades the fidelity of light projection.

\clearpage

\section*{Supplementary Note 2: Calibration}
\label{sec:calibration}
We conducted a calibration process to determine the intrinsic and extrinsic parameters of the fisheye cameras utilized in our experimental setup. To estimate the fisheye intrinsic parameters, we captured 40 images of a checkerboard at various poses and employed the MATLAB calibration toolbox\cite{scaramuzza2006toolbox, scaramuzza2006flexible, rufli2008automatic}. This allowed us to obtain the intrinsic parameters for each of the four fisheye cameras: $\Pi_\text{right-front}$, $\Pi_\text{left-front}$, $\Pi_\text{right-rear}$, and $\Pi_\text{left-rear}$.
With the intrinsic parameters in hand, we proceeded to estimate the extrinsic parameters. Specifically, we sought to determine the transformation from the right fisheye camera to the left fisheye camera, which is parameterized by the rotation angles around the $x$, $y$, and $z$ axes ($\theta_{x}$, $\theta_{y}$, $\theta_{z}$) and the translation vector ($[t_x, t_y, t_z]$). These extrinsic parameters, denoted as $\Omega=\{\theta_{x}, \theta_{y}, \theta_{z},t_x, t_y, t_z\}$, were separately defined for the frontal and rear stereo cameras as $\Omega_\text{front}$ and $\Omega_\text{rear}$, respectively.

\begin{figure}[t]
	\centering
		\includegraphics[width=\columnwidth]{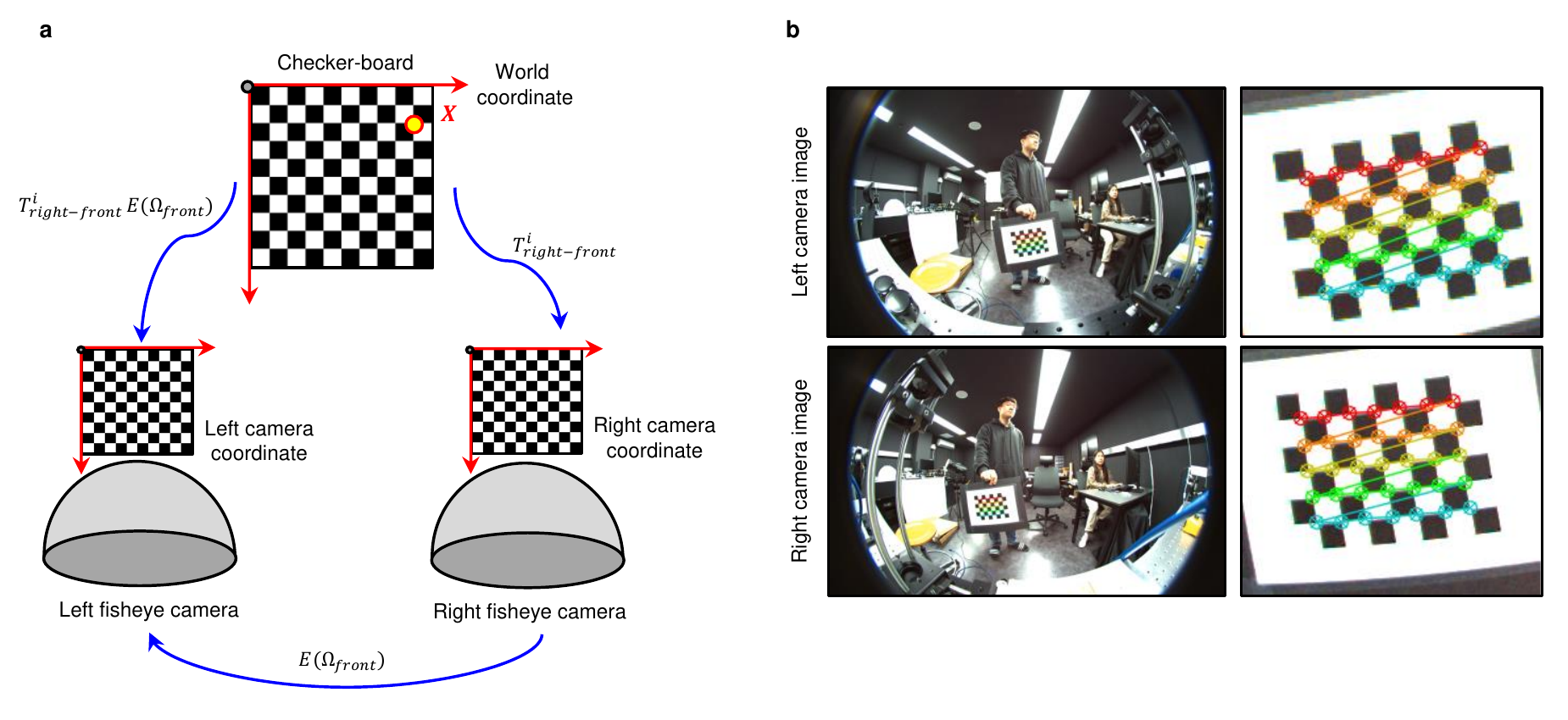}
		\caption{\textbf{a} Extrinsic calibration between the left camera and the right camera. \textbf{b} Observed corner points of the checkerboard.}
		\label{fig:calibration}
\end{figure}

To find the extrinsic parameters $\Omega_\text{front}$, we minimize the reprojection error, which represents the Euclidean distance between the projected points and the observed corner points (Fig.~\ref{fig:calibration}a).
The objective of a corresponding optimization problem is to minimize the sum of the reprojection errors for both the left and right images:
\begin{align}
 \underset{\{T_\text{right-front}^i\}_i, \Omega_\text{front}}{\text{minimize}} \, & \mathcal{L}_\text{left} + \mathcal{L}_\text{right}, \\
 \mathcal{L}_\text{right} &= \sum_{i} \|f_\text{proj}^\text{cam}(T_\text{right-front}^i X; \Pi_\text{right-front}) - p_\text{right-front}\|_2, \\
 \mathcal{L}_\text{left} &= \sum_{i} \|f_\text{proj}^\text{cam}(T_\text{right-front}^i E(\Omega_\text{front}) X;\Pi_\text{left-front}) - p_\text{left-front}\|_2, 
\end{align}
where $T_\text{right-front}^i$ represents the pose of the $i$-th checkerboard with respect to the right-front camera coordinate. $X$ corresponds to the homogeneous coordinates of the corner points on the checkerboard. $p_\text{left-front}$ and $p_\text{right-front}$ denote the positions of the observed corner pixels in the captured left and right images, respectively (Fig.~\ref{fig:calibration}b).

We employed a similar optimization process to determine the extrinsic parameters for the rear stereo camera, as shown in the following equations:
\begin{align}
 \underset{\{T_\text{right-rear}^i\}_i, \Omega_\text{rear}}{\text{minimize}} \, & \mathcal{L}_\text{left} + \mathcal{L}_\text{right}, \\
 \mathcal{L}_\text{right} &= \sum_{i} \|f_\text{proj}^\text{cam}(T_\text{right-rear}^i X; \Pi_\text{right-rear}) - p_\text{right-rear}\|_2, \\
 \mathcal{L}_\text{left} &= \sum_{i} \|f_\text{proj}^\text{cam}(T_\text{right-rear}^i E(\Omega_\text{rear}) X;\Pi_\text{left-rear}) - p_\text{left-rear}\|_2.
\end{align}
The rear-view variables are defined similarly to the front-view variables. 
To solve these optimization problems, we utilized a first-order optimizer in the PyTorch framework.

\clearpage

\section*{Supplementary Note 3: Comparative Discussion with Other Structured-light Methods}
\label{sec:comp360}
Table~\ref{table:other360} provides a comprehensive comparison of our neural 360$^\circ$ structured light with other notable structured-light methods in terms of various aspects.
\begin{itemize}
    \item First, in terms of optical elements, neural 360$^\circ$ structured light utilizes a metasurface designed and optimized for the downstream task, enabling the projection of complex patterns.
    \item Regarding the field of view (FoV), our method achieves a full 360$^\circ$ projection, allowing for complete coverage of the surrounding environment.
    \item In terms of the illumination pattern, neural 360$^\circ$ structured light learns a complex light pattern specifically tailored for the desired downstream task, such as 3D imaging. This customized pattern provides distinct matching cues and improves the accuracy of depth estimation.
    \item The design strategy of neural 360$^\circ$ structured light differs from conventional methods. Instead of heuristically designing the structured light pattern, our approach leverages the power of neural networks to optimize the metasurface phase map and achieve superior performance.
    \item We exploit both the transmission and reflection channels of metasurface by using polarization-sensitive metasurface. 
    \item Regarding 3D imaging demonstrations, neural 360$^\circ$ structured light stands out as the first method to enable 360$^\circ$ projection of complex patterns learned specifically for 3D imaging tasks. This capability improves the accuracy and robustness of depth estimation.
    \item Finally, neural 360$^\circ$ structured light can achieve high frame rates, making it suitable for real-time applications. The exact frame rate may vary depending on the specific implementation and hardware setup.
\end{itemize}

In summary, neural 360$^\circ$ structured light offers the unique capability of projecting complex patterns for 360$^\circ$ imaging and demonstrating superior performance compared to existing methods in terms of design strategy, customization, and accuracy.
\begin{table}
  \scriptsize 
  \centering
    \caption{Comparison with existing structured-light methods.}
    \hspace*{-1cm} 
    \renewcommand{\arraystretch}{3}
    \begin{tabularx}{1\linewidth}{llllllll}
    \toprule
    Method & FoV
    & Optics & Illumination pattern & Design Strategy& Polarization & \makecell[l]{3D imaging\\demonstration}\\ 
    \midrule
    \makecell[l]{Neural 360$^\circ$  \\ structured light } & 360$^\circ$ & Metasurface & \makecell[l]{Learned,\\ arbitrary} & \makecell[l]{End-to-end network\\ gradient descent} & Sensitive & Yes \\
Kim et al.\cite{kim2022metasurface} & 180$^\circ$ & Metasurface & Point clouds & \makecell[l]{GS algorithm,\\Supercell} & Insensitive & Yes \\ 
Baek et al.\cite{baek2021polka} & 90$^\circ$ & DOE & \makecell[l]{Learned,\\ arbitrary} & \makecell[l]{End-to-end network \\gradient descent} & Insensitive & Yes\\
Ni et al.\cite{Ni2020metasurface} & 120$^\circ$ & Metasurface & Point clouds & \makecell[l]{Interior point method,\\Supercell} & Insensitive & No \\
Li et al.\cite{Li2019large} & 30$^\circ$ & Metasurface & Point clouds & Supercell & Insensitive & No \\
Li et al.\cite{li2018full} & 360$^\circ$ & Metasurface & \makecell[l]{Random \\point clouds} & \makecell[l]{GS algorithm,\\ Supercell} & Insensitive & No \\
Jing et al.\cite{jing2022single} & 88.13$^\circ$ & Metasurface &  \makecell[l]{Random \\point clouds} & \makecell[l]{GS algorithm,\\ Hamming distance} & Sensitive & Yes \\  
Wang et al.\cite{Wang2021On} & 124$^\circ$ & Metasurface & Holographic image & GS algorithm & Insensitive & No \\ 

Butt et al.\cite{Butt2012Carbon} & 100$^\circ$ & Metasurface & Holographic image & GS algorithm & Insensitive & No \\ 

Chen et al.\cite{chen20202pi} & 180$^\circ$ & Metasurface & \makecell[l]{Near-uniform \\ intensity distribution} & \makecell[l]{GS algorithm,\\Rayleigh-Sommerfeld \\diffraction} & Insensitive & No \\
    \bottomrule
    \end{tabularx}%
  \label{table:other360}%
\end{table}%

\clearpage

\section*{Supplementary Note 4: Comparison with Other Wave Propagators}
\label{sec:comp_prop}
\begin{figure}[t]
	\centering
		\includegraphics[width=\columnwidth]{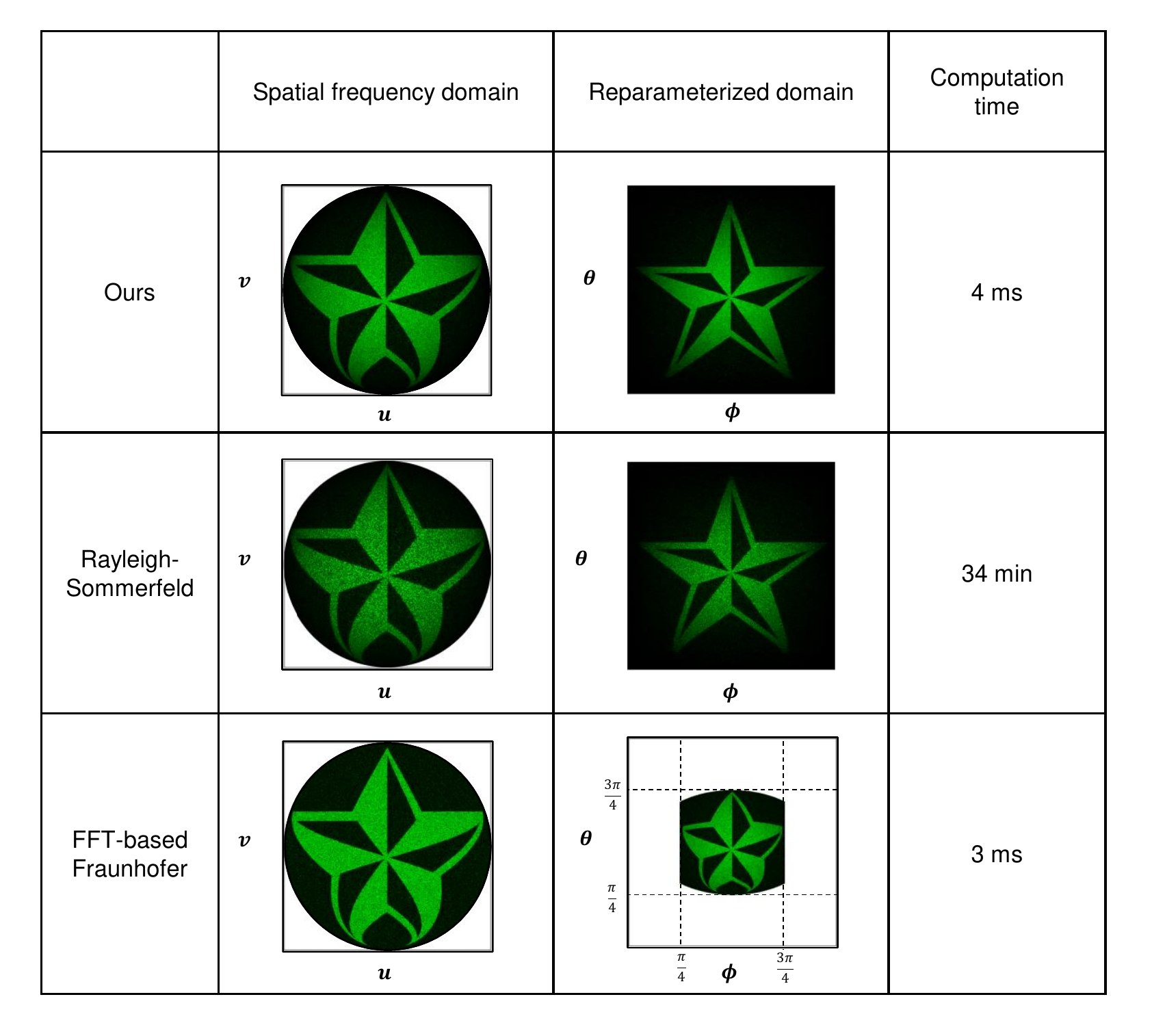}
		\caption{Calculated illumination intensity pattern and computation time taken using different propagation models.}
		\label{fig:prop_comp}
\end{figure}

Our wave propagation model enables efficient and differentiable simulation of 180$^\circ$ far-field propagated pattern that is diffracted from a metasurface (Fig.~\ref{fig:prop_comp}a).
Here, we compare our proposed model with the Rayleigh-Sommerfeld diffraction integral and the Fraunhofer diffraction integral . 
The Rayleigh-Sommerfeld diffraction integral serves as a ground-truth method that can accurately model 180$^\circ$ diffracted far-field pattern by numerically integrating the contribution from all source points on the metasurface.
However, computationally evaluating the integration results in a long computation time of 34 minutes at the resolution of a {$512 \times 512$} phase map (Fig.~\ref{fig:prop_comp}b).
For the Fraunhofer diffraction integral, we can accelerate the evaluation speed by employing FFT. 
However, due to the enforced paraxial approximation relationship, it cannot model the 180$^\circ$ angular region, with the limited angular range as $45^\circ< \theta <135^\circ$ and $45^\circ< \phi <135^\circ$ (Fig.~\ref{fig:prop_comp}c).
For evaluation, we use the metasurface phase map obtained via the phase-matching described in our main manuscript. The pixel pitch of the phase map is 260 nm, the wavelength is 532 nm, and the propagation distance $\rho$ is 1\,m.

To map the Fraunhofer diffraction results to the spherical coordinate, we obtain the closed-form expression of $\phi$ and $\theta$:
\begin{align}
    \phi &= \cot^{-1}{\left(u\lambda \right)}, \\
    \theta &= \cot^{-1} \left(v\sin{\phi}\lambda\right).
\end{align}
The derivation for obtaining such close-form expression is to use the paraxial approximation in Fraunhofer diffraction that assumes the following relationship for the spatial frequencies:
\begin{align}
    u &= \frac{x}{z\lambda}, \\
    v &= \frac{y}{z\lambda}.
\end{align}
We also use the coordinate conversion form the spatial frequencies to the spherical coordinate:
\begin{align}
        u &= \frac{x}{z\lambda} = \frac{\rho\sin{\theta}\cos{\phi}}{\rho\sin{\theta}\cos{\phi}\lambda} = \frac{\cot{\phi}}{\lambda}, \\
    v &= \frac{y}{z\lambda} = \frac{\rho\cos{\theta}}{\rho\sin{\theta}\sin{\phi}\lambda} = \frac{\cot{\theta}}{\sin{\phi}\lambda}.
\end{align}

\clearpage

\section*{Supplementary Note 5: Derivation of the Proposed Propagation Model}
\label{sec:derivation}
\begin{figure}[t]
	\centering
		\includegraphics[width=\columnwidth]{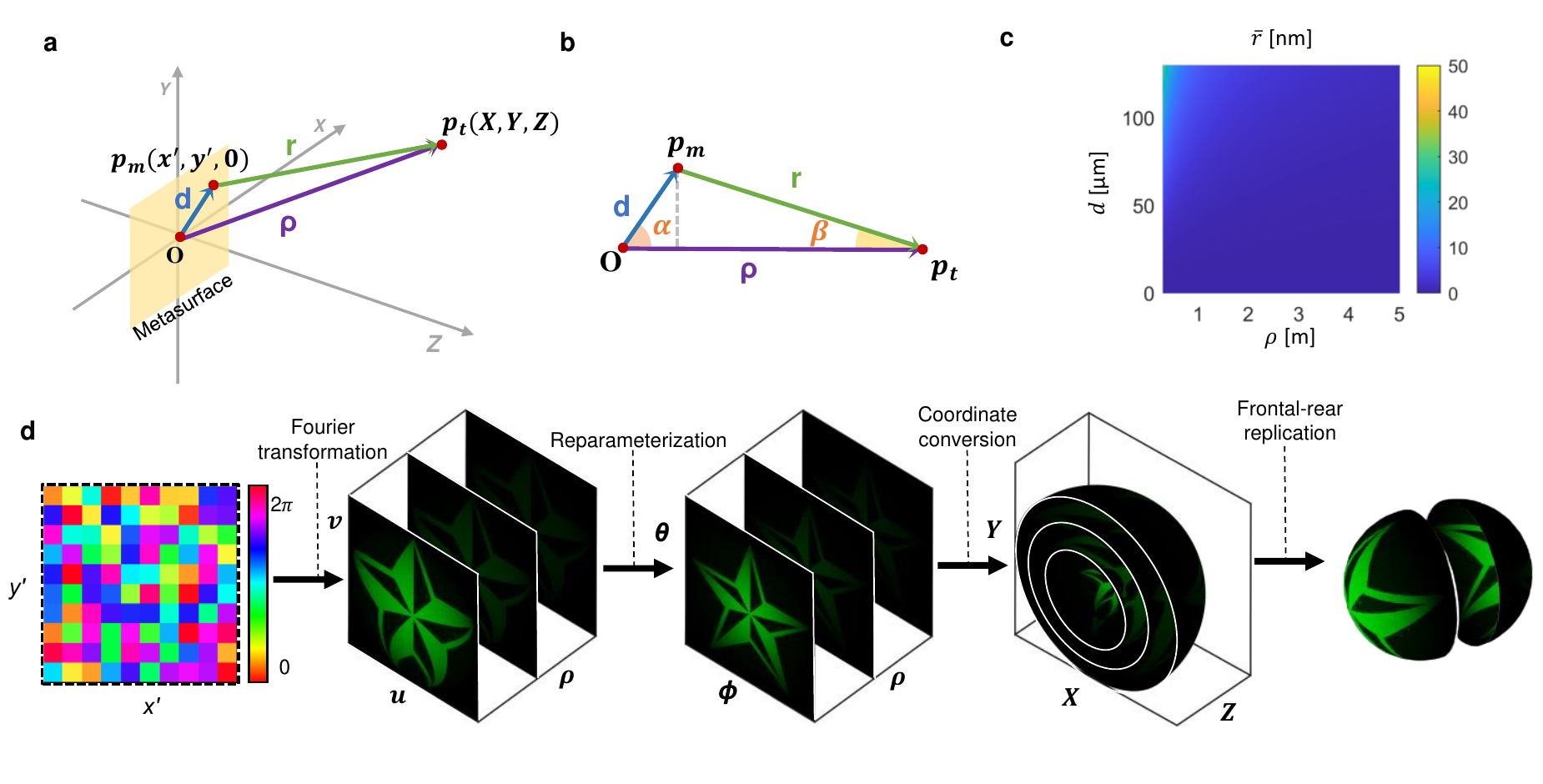}
\caption{\textbf{a} Geometric relationship in the Cartesian coordinate of the metasurface, the source point, and the propagated target point. \textbf{b} The angle $\alpha$ is defined by the two vectors $\overrightarrow{O p_\text{m}}$ and $\overrightarrow{O p_\text{t}}$. The angle $\beta$ is defined by the two vectors $\overrightarrow{p_\text{t} p_\text{m}}$ and $\overrightarrow{p_\text{t}O}$. \textbf{c} $r_{error, max}$  plot with respect to $d$ and $\rho$. \textbf{d} Overview of the computational process of our propagation model.}
		\label{fig:suppl_derivation}
\end{figure}

This section describes the derivation and validity of our propagation model. We assume the center of the metasurface is aligned with the origin $O$ (Fig.~\ref{fig:suppl_derivation}a). $\rho$ represents the distance between a target 3D point $p_t$ and the origin $O$. $\theta$ is the angle deviating from the y-axis and $\phi$ is the angle deviating from the x-axis. $r$ is the distance between the source point $p_{m}$ and the target point $p_{t}$ and $d$ is the distance between the source point $p_s$ and $O$. 
We begin with the Rayleigh-Sommerfeld propagation 
\begin{equation}
\label{eq:RS_equation}
 U(x, y, z) = \frac{1}{\lambda j} \int\int U'(x', y', 0)\frac{ze^{jkr}}{r^2} dx'dy'.
\end{equation}
$k=\frac{2\pi}{\lambda}$ is the wave number.
Using the defined spherical coordinate parameters, $\rho$, $\theta$ and $\phi$, the target point $p_{t}$ in Cartesian coordinate, $X$ and $Y$ and $Z$, can be expressed as, 
\begin{equation}
\label{eq:spherical_coord}
p_{t} = (X, Y, Z) = \rho (\sin{\theta}\cos{\phi}, \cos{\theta}, \sin{\theta}\sin{\phi}).
\end{equation}
Plugging the above into eq(\ref{eq:RS_equation}), we obtain the following equation expressed in spherical coordinate:
\begin{equation}
\label{eq:RS_spherical}
 U(\rho, \theta, \phi) = \frac{1}{\lambda j} \int\int U'(x', y', 0)\frac{e^{jkr}}{r^2}\rho\sin{\theta}\sin{\phi}\, dx'dy'.
\end{equation}
We then write the distance $\rho$ between the target point $p_t$ and the origin $O$ as 
\begin{equation}
    \rho = d\cos{\alpha}+r\cos{\beta},
\end{equation}
where $\alpha$ and $\beta$ are the angles formed by the vectors $\overrightarrow{Op_m}$ and $\overrightarrow{Op_t}$, and $\overrightarrow{p_t p_m}$ and $\overrightarrow{p_t O}$ respectively.
We assume $d \ll \rho$.
This is commonly valid for the far-field propagation of metasurface-modulated wavefront where the propagation distance $\rho$ is sufficiently larger than the metasurface radius $d$.  
Applying the approximation, we rewrite the distance $r$ as
\begin{equation}
    {r}=\rho-d \cos\alpha \approx r,
\end{equation}
as the value of $\cos {\beta}$ approaches to one, given $d \ll \rho$.
We then rewrite the distance ${r}$ as 
\begin{equation}
    {r}=\rho \left( 1 - \frac{x'X+y'Y}{\rho^2} \right),
\end{equation}
using the dot product of two vectors: $\overrightarrow{Op_s} \cdot \overrightarrow{Op_t}= x'X+y'Y = d\rho \cos {\alpha}$.
Since $\frac{d \cos \alpha}{\rho} \approx 0$, we obtain ${r}^2\approx \rho^2$.
Applying this to eq(\ref{eq:RS_spherical}), we arrive at
\begin{align}
 U(\rho, \theta, \phi) &= \frac{\sin{\theta}\sin{\phi}}{ \rho\lambda j} \int\int U'(x', y', 0){e^{jkr}}\, dx'dy'\\
&= \frac{e^{jk\rho}\sin{\theta}\sin{\phi}}{\rho\lambda j} \int\int U'(x', y', 0) e^{-2\pi j(\frac{X}{\lambda}x' + \frac{Y}{\lambda})y'} dx'dy'\\
&= \frac{e^{jk\rho}\sin{\theta}\sin{\phi}}{\rho\lambda j} \int\int U'(x', y', 0) e^{-2\pi j(\frac{\sin{\theta}\cos{\phi}}{\lambda}x' + \frac{\cos{\theta}}{\lambda})y'} dx'dy'. \label{eq:RS_approx}
\end{align}
Eq(\ref{eq:RS_approx}) can be interpreted as a Fourier transform with spatial frequencies, $u = \frac{\sin{\theta}\cos{\phi}}{\lambda}$ and $v = \frac{\cos{\theta}}{\lambda}$. Using these relations, we can map each discrete Fourier transformed output value into a physical location with corresponding spherical coordinate ($\theta$, $\phi$). (Fig.\ref{fig:suppl_derivation}b).

\clearpage

\section*{Supplementary Note 6: Applicable Range of the Proposed Propagation Model}
\label{sec:validity}
Here, we discuss the applicable range of the proposed propagation model.
In the derivation of our propagation model, we approximated the distance $r$ as 
\begin{align}
r &= \sqrt{{\rho}^2 + d^2 - 2d\rho\cos{\alpha}}\\
 &\approx \rho - d\cos{\alpha} = \hat{r},
\end{align}
where $\hat{r}$ denotes the approximated distance.
See Fig.\ref{fig:suppl_derivation}c.
We then describe the error caused by the approximation as $\bar{r}$:
\begin{align}
    \bar{r} &= r - \hat{r} \\
    &=\sqrt{{\rho}^2 + d^2 - 2d\rho\cos{\alpha}} - (\rho - d\cos{\alpha}).
\end{align}
Note that the error $\bar{r}$ is always positive since $r\geq \hat{r}$.
For our propagation model being applicable, the error $\bar{r}$ should be sufficiently small compared to the wavelength $\lambda$:
\begin{equation}
\label{eq:error_cond}
    \bar{r}\ll \lambda.
\end{equation}
Thus, we find the maximum value of $\bar{r}$ with respect to $\alpha$ considering $\rho$ and $d$ as constants. 
If $\alpha$ is either $0$ or $\pi$ at extremes, the error $\bar{r}$ becomes zero. 
If $\alpha$ is not at extremes, we can obtain the maximum value of $\bar{r}$ by differentiating the error $\bar{r}$ with respect to $\alpha$ and finding the value $\alpha$ that makes the gradient zero:
\begin{gather}
\frac{\partial \bar{r}}{ \partial \alpha} = \frac{d\rho\sin{\alpha}}{\sqrt{\rho^2 + d^2 -2d\rho\cos{\alpha}}} - d\sin{\alpha} = 0.
\end{gather}
Rearranging this results in the estimate of $\alpha_\text{max}= \arccos{\frac{d}{2\rho}}$ that gives the maximum error $\bar{r}_\text{max}$:
\begin{align}
\bar{r}_\text{max} &= \sqrt{\rho^2 + d^2 -2d\rho(\frac{d}{2\rho})} - \rho + d(\frac{d}{2\rho}) \\
&= \frac{d^2}{2\rho}.\label{eq:max_error}
\end{align}
Eq(\ref{eq:max_error}) states that the upper bound of the error is determined by the metasurface radius $d$ and the travel distance $\rho$.
Note that our propagation model is applicable if the maximum error $\bar{r}_\text{max}$ is sufficiently smaller than the wavelength $\lambda$ as in eq(\ref{eq:error_cond}).
We show the value of $\bar{r}_\text{max}$ for the range of metasurface radius $d < 200\,\mu m$ and the travel distance $0.3\,m < \rho < 5\,m$ in Fig.\ref{fig:suppl_derivation}d that exhibiting smaller values than the wavelength $\lambda=550\,nm$. Thus, we use the proposed propagation model in our experiment.

\clearpage

\label{sec:light_projection}
\begin{figure}[t]
	\centering
		\includegraphics[width=\columnwidth]{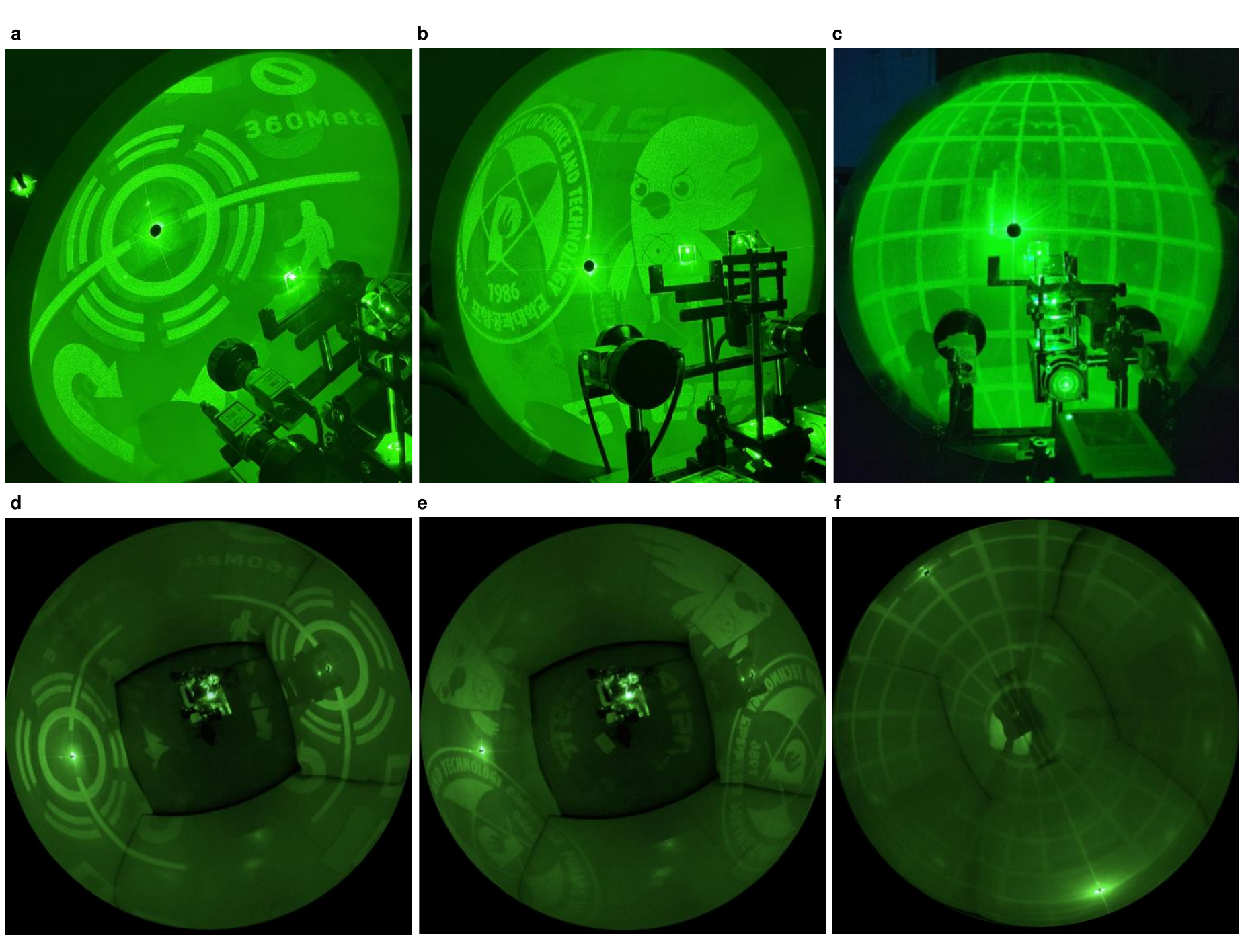}
		\caption{Additional experimental demonstration of 360$^\circ$ light projection. \textbf{a-c} 360$^\circ$ projection onto a hemispherical screen. \textbf{d-f} Bird-eye view of 360$^\circ$ projection in an enclosed room}
		\label{fig:hologram}
\end{figure}
\clearpage

\section*{Supplementary Note 7: Additional Details on Image Formation for 3D Imaging}
\label{sec:image_formation}
\subsection{Computational model for fish-eye cameras.}
We adopt the omni-directional fish-eye camera model\cite{urban2015improved, scaramuzza2006flexible}, which allows for the projection of a 3D scene point $X$ onto a camera pixel $p$ and its reverse process as unprojection (Fig.~\ref{fig:image_formation}a).
The camera model is defined by a set of parameters $\Pi=\{ A, B, \Gamma \}$, which includes the affine matrix $A$, the polynomial coefficients for projection $B=\{\beta_{0}, \beta_{1}, ..., \beta_{n}\}$ and the inverse polynomial coefficients for unprojection $\Gamma=\{\gamma_{0}, \gamma_{1}, ..., \gamma_{n}\}$ .
Using the parameter $\Pi$, we can perform projection and unprojection operations with the functions $f_\text{proj}^\text{cam}$ and $f_\text{unproj}^\text{cam}$, respectively:
\begin{align}
p &= f_\text{proj}^\text{cam}(X; \Pi), \\
X &= f_\text{unproj}^\text{cam}(p, z; \Pi),
\end{align}
where $z$ represents the depth of the pixel $p$.
These projection and unprojection functions are differentiable with respect to the camera parameters $\Pi$, allowing for the integration of the camera model into our differentiable image formation.

\subsection{Differentiable fish-eye image formation.}
\begin{figure}[t]
	\centering
		\includegraphics[width=\columnwidth]{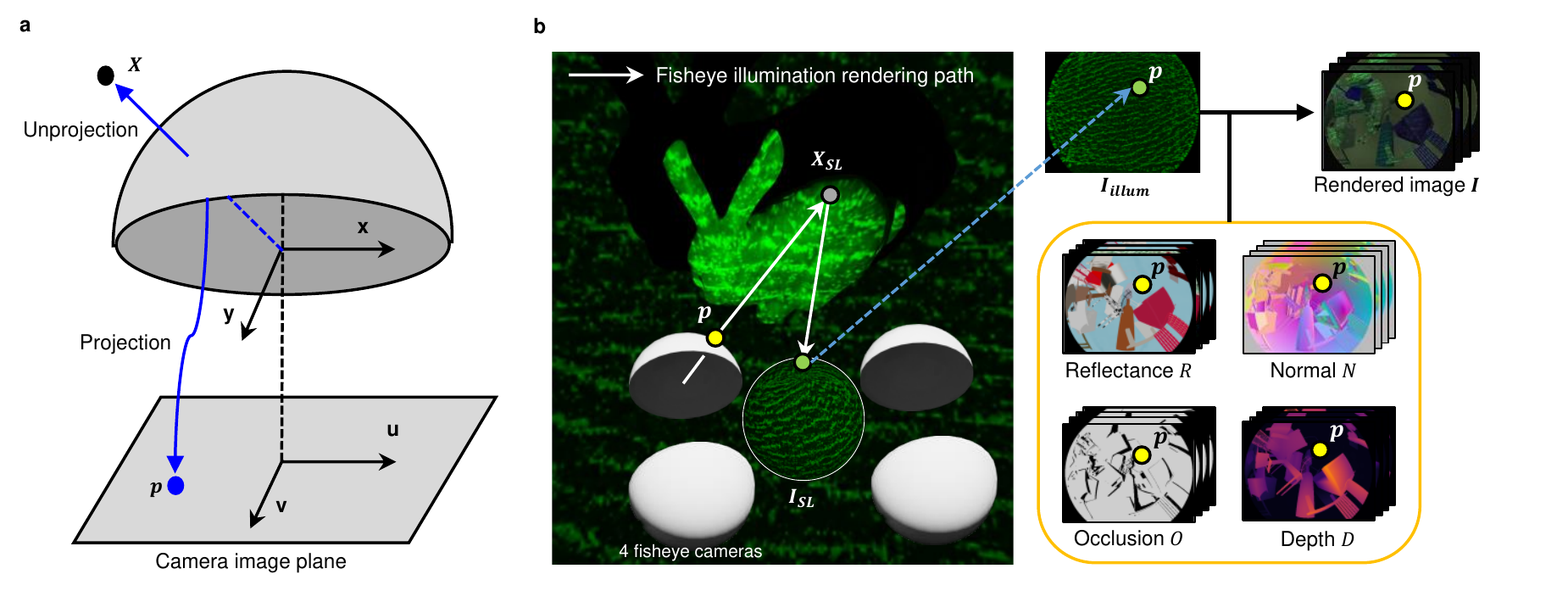}
		\caption{\textbf{a} Diagram of our fisheye camera model. $X$ represents a point in the scene and $p$ represents a pixel on the camera image plane. Blue arrows indicate the projection and unprojection path. \textbf{b} Diagram illustrating our differentiable image formation process.}
		\label{fig:image_formation}
\end{figure}

In our image formation, we simulate the rendering of a fish-eye image under neural 360$^\circ$ structured light. Fig.~\ref{fig:image_formation}b provides an overview of the image formation pipeline. Here, we describe the formulation of the rendered intensity for a camera pixel $p$, which can be generalized to all camera pixels.

First, we obtain the unprojected scene point $X_\text{SL}$ in the structured-light coordinate using the depth value $z$ associated with the camera pixel $p$:
\begin{equation}
X_\text{SL} = T_\text{cam2SL} [ f_\text{unproj}^\text{cam}(p, z ; \Pi), 1 ]^\intercal,
\end{equation}
Here, $T_\text{cam2SL}$ represents the transformation matrix from the camera coordinate to the structured-light coordinate. The depth value $z$ is known from the corresponding ground-truth depth map in our synthetic 360$^\circ$ dataset.
Next, we project the scene point $X_\text{SL}$ onto the structured-light image and sample the intensity value of the corresponding pixel:
\begin{equation}
I_\text{illum}=I_\text{SL}\left(f_\text{proj}^\text{SL}(X_\text{SL})\right),
\end{equation}
where $I_\text{SL}$ denotes the structured-light image generated by a metasurface with the phase map $\Phi$: $I_\text{SL}=\alpha f_\text{prop}(\Phi)$. The function $f_\text{proj}^\text{SL}$ represents the projection of a 3D point onto the unit sphere in the structured-light coordinate. $I_\text{illum}$ corresponds to the intensity of the structured light that illuminates the scene point corresponding to the camera pixel $p$.
Finally, the rendered intensity at the fish-eye camera pixel $p$ is computed as follows:
\begin{equation}
I(p) = f_\text{clip}\left(S(p) \odot R(p) \odot \left( O(p) \odot I_\text{illum} + \beta \right) + \eta \right),
\end{equation}
where $S(p)$ represents the cosine foreshortening term, $R(p)$ is the scene reflectance, and $O(p)$ denotes the occlusion map indicating whether the structured light is visible at the camera pixel $p$.
This differentiable image formation process allows us to simulate the rendering of a fish-eye image under neural 360$^\circ$ structured light, incorporating the effects of depth, reflectance, occlusion, and other parameters, while accounting for noise and clipping.

\clearpage

\section*{Supplementary Note 8: 360$^\circ$ Depth-estimation Neural Network}
\label{sec:360_depth}
We develop a neural 360$^\circ$ depth reconstruction method that estimates 180$^\circ$ frontal and rear depth maps, resulting in a complete 360$^\circ$ depth map. 
Our depth-reconstruction method is differentiable and exploits the information of neural 360$^\circ$ structured light encoded in the captured images.The depth reconstruction network, as illustrated in Fig.~\ref{fig:reconstructor}b, comprises several components: a feature extractor $f_\text{feat}$, spherical sweeping $f_\text{sweep}$, cost aggregator $f_\text{cost}$, and edge-aware upsampler $f_\text{edge}$.
Here, we describe the depth estimation for the frontal view. Its extension to the rear view can be applied in the same manner.

\begin{enumerate}
    \item The feature extractor $f_\text{feat}$ employs convolutional and downsampling layers to extract feature maps $F_i$ from the captured images $I_i$:
\begin{align}
F_\text{right-front} &= f_\text{feat}(I_\text{right-front}), \\
F_\text{left-front} &= f_\text{feat}(I_\text{left-front}).
\end{align}
The feature extractor reduces the spatial resolution of the raw images, providing useful features for depth estimation while reducing computational costs.
    Table.~\ref{table:feature} shows the detailed network architecture of the feature extractor.  
    \item The feature maps extracted from the left and right images are processed using the spherical sweeping function $f_\text{sweep}$. Spherical sweeping involves unprojecting a right-camera pixel $p$ to obtain a scene point $X$ with a specific depth $z_i \in \{z_1, z_2, \cdots, z_{90}\}$, and then projecting this scene point onto the left camera to obtain the corresponding pixel $q$. The cost $C_\text{front}(p,z_i)$ is computed by comparing the corresponding feature maps $F_\text{right-front}(p)$ and $F_\text{left-front}(q)$:
\begin{equation}
C_\text{front}(p,z_i) =\| F_\text{right-front}(p) - F_\text{left-front}(q) \|_2.
\end{equation}
The spherical sweeping function generates a cost volume $C\text{front}$, representing the matching costs between pixels and depth markers.
We uniformly sample the depth candidates $\{z_1, z_2, \cdots, z_{90}\}$ in an inverse depth from 0.3\,m to 5\,m.
    In summary, the spherical sweeping function takes the extracted features as inputs and outputs the cost volume:
    \begin{equation}
        C_\text{front}=f_\text{sweep}(F_\text{right-front}, F_\text{left-front}).
    \end{equation}
    
    \item 
    The cost aggregator $f_\text{cost}$ filters the cost volume $C_\text{front}$ and estimates the depth map by aggregating the matching costs. The aggregator employs 3D convolutional layers and a softmax function to transform the cost volume into a probability map. The estimated depth map $\dot{D}_\text{right-front}$ is computed by summing the products of each probability and depth candidate value:
        \begin{equation}
        \dot{D}_\text{right-front}=f_\text{cost}(C_\text{front}).
    \end{equation}
Table.~\ref{table:filter} shows the network architecture.

    \item 
    To upsample the estimated depth map while preserving edge details, we introduce an edge-aware upsampler $f_\text{edge}$ that utilizes the input image $I_\text{right-front}$ as a guide. The upsampler enhances the resolution of the depth map, resulting in the high-resolution depth map $\hat{D}_\text{right-front}$:
            \begin{equation}
        \hat{D}_\text{right-front}=f_\text{edge}(\dot{D}_\text{right-front}, I_\text{right-front}).
    \end{equation}
See Table.~\ref{table:edge} for the network architecture of the edge-aware upsampler.

\end{enumerate}

\subsection{Occlusion filtering.}
\begin{figure}[t]
	\centering
		\includegraphics[width=\columnwidth]{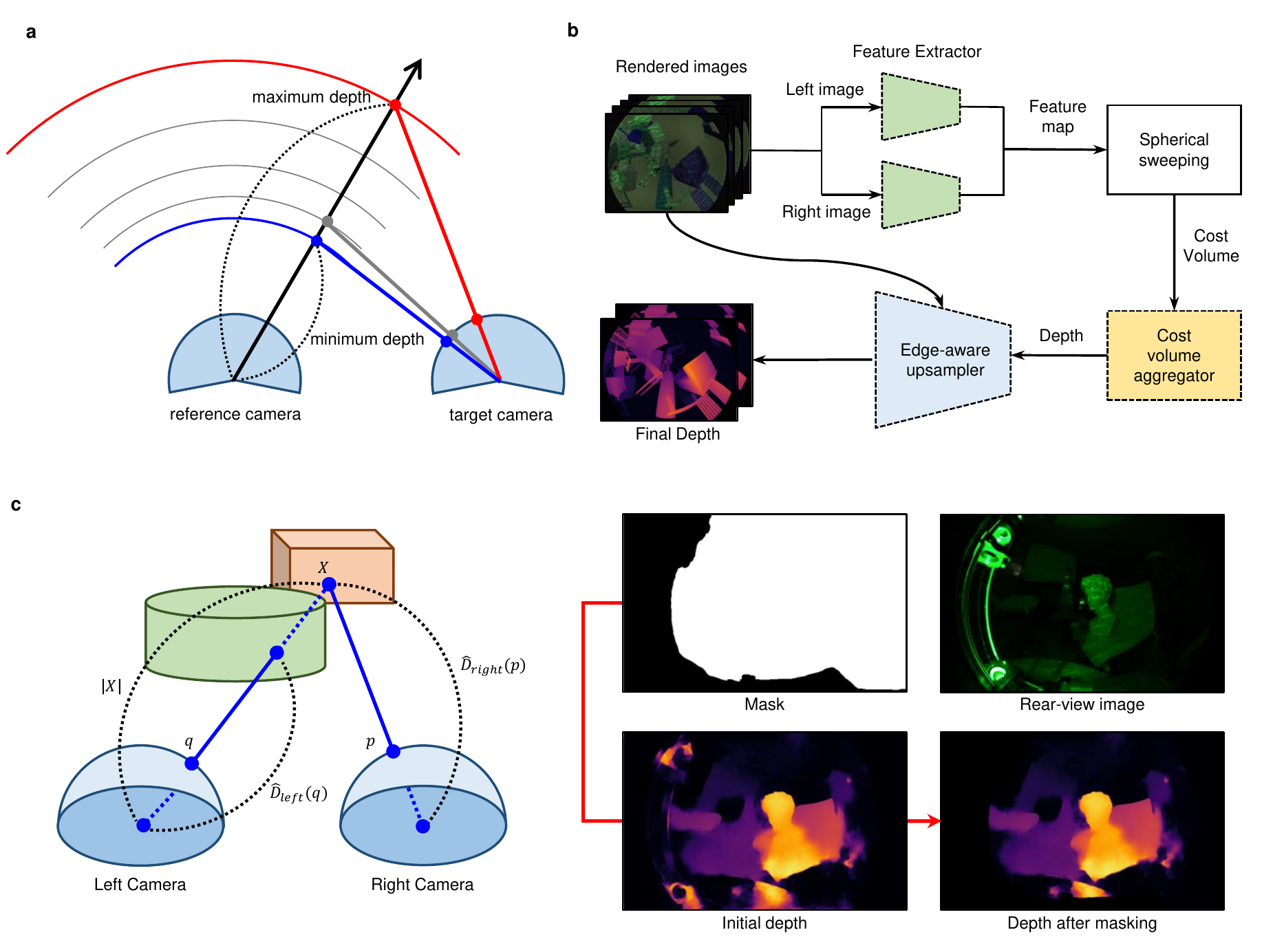}
		\caption{\textbf{a} Diagram of the spherical sweeping process between stereo cameras. The blue point represents the unprojected scene point by a minimum depth value. The red point represents the unprojected scene point by a maximum depth value. \textbf{b} Overview of depth reconstructor for 3D imaging. Three modules marked with dotted lines are neural networks. \textbf{c} Left-right consistency check for occlusion detection. \textbf{d} Masking the illumination module in the rear view.}
		\label{fig:reconstructor}
\end{figure}

Geometric occlusion between stereo images presents a challenge for correspondence matching, as occluded scene points are only visible in one of the cameras (Fig.~\ref{fig:reconstructor}). To address this issue, we employ a left-right consistency check\cite{godard2017unsupervised} to detect and filter out occluded pixels in real-world captured images.
First, we estimate the left-view and right-view depth maps, $\hat{D}_\text{left-front}$ and $\hat{D}_\text{right-front}$, respectively, using our depth reconstructor.
For each right-camera pixel $p$, we unproject it to obtain the corresponding scene point $X$ using the estimated right-camera depth $\hat{D}_\text{right}(p)$. Next, we project this scene point onto the left camera to obtain the corresponding pixel $q$ and retrieve the estimated left-camera depth $\hat{D}_\text{left}(q)$.
To determine if the right-camera pixel $p$ is occluded, we compare the left-camera depth with the depth of the corresponding scene point from the right-camera pixel. Specifically, we apply the following occlusion check:
\begin{align}
\text{$p$ is occluded } \, &\text{, if } \frac{1}{\left| \hat{D}_\text{left}(q) - \|X\|_2 \right|} < \delta, \\
\text{$p$ is not occluded} \, &\text{, otherwise},
\end{align}
where $\delta$ is a threshold parameter.
By performing the left-right consistency check, we can identify and filter out occluded pixels.
Also, in our experimental setup, the rear-view cameras inadvertently capture the illumination module due to its placement (Fig.~\ref{fig:reconstructor}d). To ensure accurate depth map visualization of real-world scenes, we apply post-processing steps to remove the occluded pixels detected through the left-right consistency check and manually segment the captured illumination module.

\begin{table}
  \scriptsize 
  \centering
    \caption{Network architecture of feature extractor}
    \hspace*{-1cm} 
    \renewcommand{\arraystretch}{2}
    \begin{tabularx}{1\linewidth}{lllllllll}
    \toprule
    Layer Name & Layer Type & Input Channel & Output Channel & Kernel & Stride & Padding & Dilation & Activation\\
    \midrule
    downsample & conv2d & 3 & 34 & 2x2 & 2 & 2 & 1 \\ 
    downsample & conv2d & 34 & 34 & 2x2 & 2 & 2 & 1\\ 
    res-block & conv2d & 34 & 34 & 3x3 & 1 & 1 & 1 & LReLU \\ 
    res-block & conv2d & 34 & 34 & 3x3 & 1 & 1 & 1 & LReLU \\ 
    res-block & conv2d & 34 & 34 & 3x3 & 1 & 1 & 1 & LReLU \\ 
    res-block & conv2d & 34 & 34 & 3x3 & 1 & 1 & 1 & LReLU \\ 
    res-block & conv2d & 34 & 34 & 3x3 & 1 & 1 & 1 & LReLU \\ 
    res-block & conv2d & 34 & 34 & 3x3 & 1 & 1 & 1 & LReLU \\ 
    conv-alone & conv2d & 34 & 34 & 3x3 & 1& 1 & 1 & LReLU \\ 
    \bottomrule
    \end{tabularx}%
  \label{table:feature}%
\end{table}%

\begin{table}
  \scriptsize 
  \centering
    \caption{Network architecture of cost volume aggregator}
    \hspace*{-1cm} 
    \renewcommand{\arraystretch}{2}
    \begin{tabularx}{1\linewidth}{lllllllll}
    \toprule
    Layer Name & Layer Type & Input Channel & Output Channel & Kernel & Stride & Padding & Dilation & Activation\\
    \midrule
    conv3d & conv3d & 34 & 34 & 3x3 & 1 & 1 & 1 & LReLU \\ 
    conv3d & conv3d & 34 & 34 & 3x3 & 1 & 1 & 1 & LReLU \\ 
    conv3d & conv3d & 34 & 34 & 3x3 & 1 & 1 & 1 & LReLU \\ 
    conv3d & conv3d & 34 & 34 & 3x3 & 1 & 1 & 1 & LReLU \\ 
    conv3d & conv3d & 34 & 34 & 3x3 & 1 & 1 & 1 & LReLU \\ 
    conv3d & conv3d & 34 & 34 & 3x3 & 1 & 1 & 1 & LReLU \\ 
    conv3d & conv3d & 34 & 34 & 3x3 & 1 & 1 & 1 & LReLU \\ 
    conv3d-alone & conv3d & 34 & 1 & 3x3 & 1 & 1 & 1 \\ 
    \bottomrule
    \end{tabularx}%
  \label{table:filter}%
\end{table}%

\begin{table}
  \scriptsize 
  \centering
    \caption{Network architecture of edge-aware upsampler}
    \hspace*{-1cm} 
    \renewcommand{\arraystretch}{2}
    \begin{tabularx}{1\linewidth}{lllllllll}
    \toprule
    Layer Name & Layer Type & Input Channel & Output Channel & Kernel & Stride & Padding & Dilation & Activation\\
    \midrule
    edge-in & conv2d & 4 & 32 & 3x3 & 1 & 1 & 1 & BN-LReLU \\ 
    edge-filter & conv2d& 32 & 32 & 3x3 & 1 & 1 & 1 & LReLU \\ 
    edge-filter & conv2d& 32 & 32 & 3x3 & 1 & 1 & 1 & LReLU \\ 
    edge-filter & conv2d& 32 & 32 & 3x3 & 1 & 1 & 1 & LReLU \\ 
    edge-filter & conv2d& 32 & 32 & 3x3 & 1 & 1 & 1 & LReLU \\ 
    edge-filter & conv2d& 32 & 32 & 3x3 & 1 & 1 & 1 & LReLU \\ 
    edge-out & conv2d & 32 & 1 & 3x3 & 1 & 1 & 1 \\ 
    conv2d-feature & conv2d & 2 & 32 & 3x3 & 1 & 1 & 1 & LReLU \\ 
    res-astrous & conv2d & 32 & 32 & 3x3 & 1 & 1 & 1 & LReLU \\ 
    res-astrous & conv2d & 32 & 32 & 3x3 & 1 & 1 & 2 & LReLU \\ 
    res-astrous & conv2d & 32 & 32 & 3x3 & 1 & 1 & 3 & LReLU \\ 
    res-astrous & conv2d & 32 & 32 & 3x3 & 1 & 1 & 4 & LReLU \\ 
    res-astrous & conv2d & 32 & 32 & 3x3 & 1 & 1 & 3 & LReLU \\ 
    res-astrous & conv2d & 32 & 32 & 3x3 & 1 & 1 & 2 & LReLU \\ 
    res-astrous & conv2d & 32 & 32 & 3x3 & 1 & 1 & 1 & LReLU \\ 
    conv2d-out & conv2d & 32 & 1 & 3x3 & 1 & 1 & 1 & ReLU \\ 
    \bottomrule
    \end{tabularx}%
  \label{table:edge}%
\end{table}%
\clearpage

\section*{Supplementary Note 9: Additional Details on End-to-end Optimization for 3D Imaging}
\label{sec:echo}
Our approach to end-to-end optimization for 3D imaging involves two training stages. In the first stage, we simultaneously optimize the metasurface phase map $\Phi$ and the network parameters of the depth reconstructor $\Theta$. The loss function $\mathcal{L}_\text{1}$ in this stage is composed of multiple components:
\begin{equation}
\mathcal{L}_\text{1} = \lambda_\text{MSE} \mathcal{L}_\text{MSE}(\hat{D}, D) + \lambda_\text{TV} \mathcal{L}_\text{TV} (\hat{D}) + \lambda_\text{SL} \mathcal{L}_\text{SL}(\alpha|f_\text{prop}(\Phi)|),
\end{equation}
where $\mathcal{L}_\text{MSE}$ is the mean-squared-error loss between the reconstructed 360$^\circ$ depth $\hat{D}$ and the ground-truth 360$^\circ$ depth $D$. The term $\mathcal{L}_\text{TV}$ represents the total variation loss of the reconstructed depth, while $\mathcal{L}_\text{SL}$ serves to prevent energy concentration on a few structured-light points. Specifically, $\mathcal{L}_\text{SL}(S) = \frac{1}{\|\nabla S\|1}$, where $S$ is the structured-light image $S = \alpha|f\text{prop}(\Phi)|$. The weights $\lambda_\text{MSE}$, $\lambda_\text{TV}$, and $\lambda_\text{SL}$ are set to $1.0$, $0.4$, and $0.08$.

\begin{figure}[t]
	\centering
		\includegraphics[width=\columnwidth]{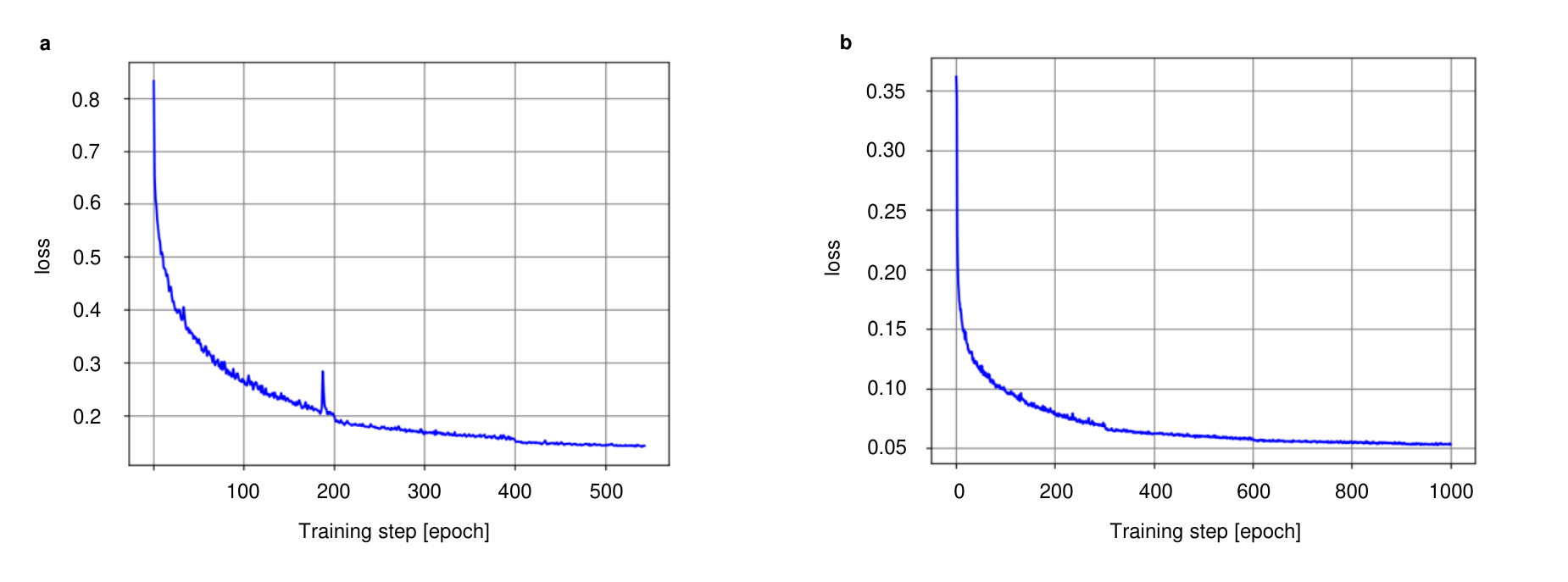}
		\caption{\textbf{a} Training loss graph of the first stage. \textbf{b} Training loss graph of the second stage.}
		\label{fig:loss}
\end{figure}

Once the training of the first stage reaches convergence and the shape of the structured-light pattern stabilizes, we transition to the second stage. In this stage, we keep the phase map of the metasurface fixed and continue optimizing the reconstructor using the following loss function:
\begin{equation}
\mathcal{L}_\text{2} = \lambda_\text{MSE} \mathcal{L}_\text{MSE}(\hat{D}, D) + \lambda_\text{TV} \mathcal{L}_\text{TV} (\hat{D}).
\end{equation}
This two-stage training approach allows for effective training of the depth reconstructor, as demonstrated by the loss graph in Fig.~\ref{fig:loss}.

To train each stage, we utilized a single NVIDIA A6000 GPU for approximately 20 hours, with a batch size of 4. The optimization of the phase map was performed with a learning rate of $10^{-2}$, while the reconstructor was optimized with a learning rate of $3 \times 10^{-3}$ using the Adam optimizer\cite{kingma2014adam}. Additionally, we use the learning rate decay every 350 steps using an LR scheduler in PyTorch. The phase distributions of the metasurface were initialized from a uniform random distribution.

\clearpage

\section*{Supplementary Note 10: Experimental Comparison of 3D Imaging with Multi-dot 360$^\circ$ Structured Light and Passive Stereo}
\label{sec:comparison}
We evaluate the performance of 3D imaging using neural 360$^\circ$ structured light against 360$^\circ$ multi-dot structured light and passive stereo. The multi-dot structured light, based on a heuristic metasurface design derived from the supercell method\cite{kim2022metasurface}, was fabricated into a metasurface and then incorporated into our illumination module for testing. We maintained consistent settings of the illumination modules, including laser power and camera exposure when using the multi-dot structured light. We trained the same network architecture as the depth reconstructor for each case with their respective data: with the multi-dot structured light and without structured light under {sufficient ambient light conditions}.

\subsection{Qualitative evaluation.}
\begin{figure}[t]
	\centering
		\includegraphics[width=\columnwidth]{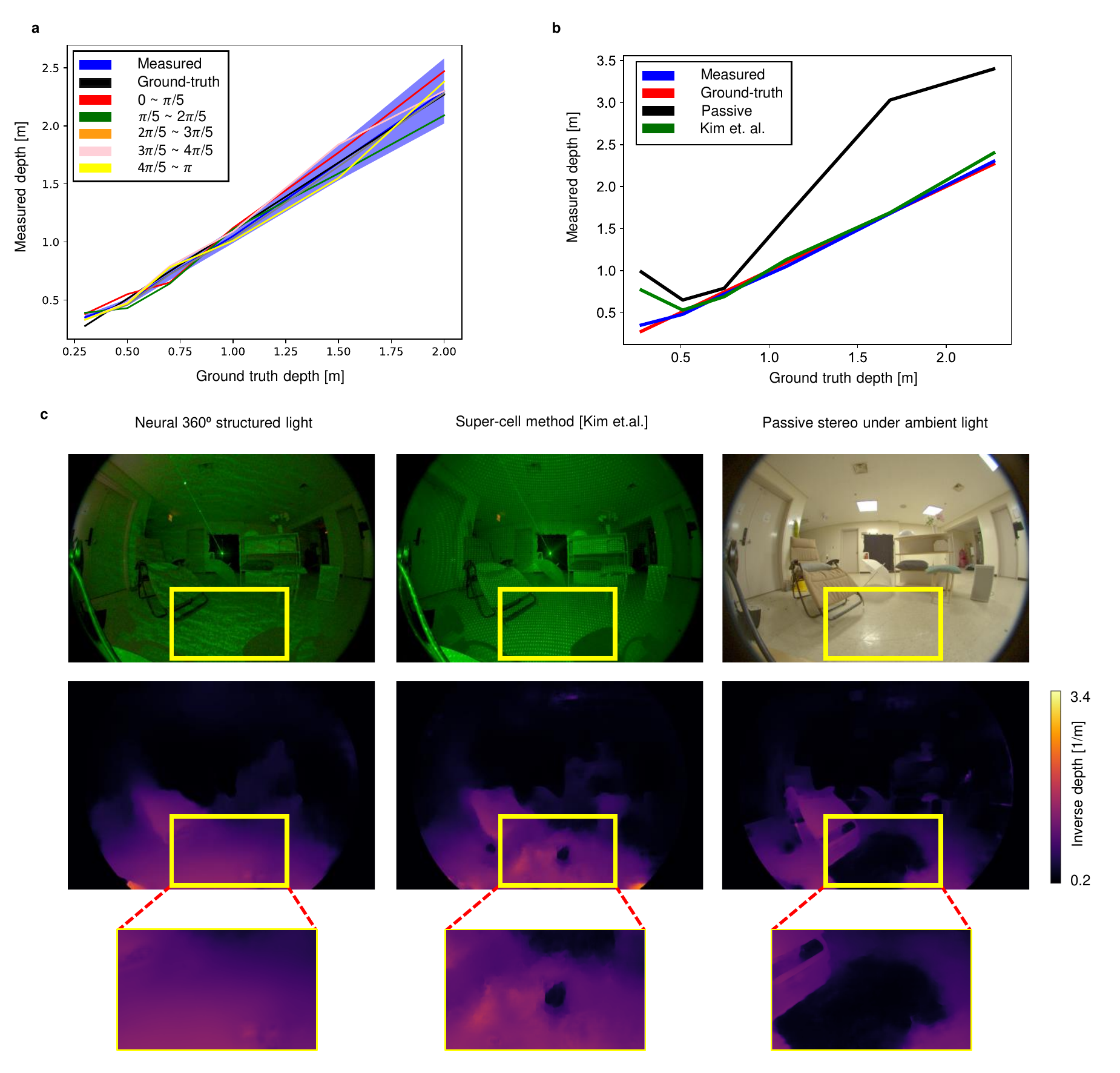}
		\caption{Qualitative comparison of depth estimation. The yellow box in the images corresponds to a flat and smooth floor. Neural 360$^\circ$ structured light exploits the learned non-uniform features for robust 3D imaging while the super-cell method\cite{kim2022metasurface} and passive stereo struggle in recovering accurate depth.}
		\label{fig:depth_accuracy}
\end{figure}

We captured an indoor scene with various objects as shown in Fig.\ref{fig:depth_accuracy}.
The yellow boxes in the image represent a flat and smooth floor. Neural 360$^\circ$ structured light successfully reconstructs the depth map of the scene, while the multi-dot structured light and passive method show noticeable holes and lack smoothness in their results.
When compared to the heuristically-designed multi-dot structured light\cite{kim2022metasurface}, our neural 360$^\circ$ structured light with learned metasurface yields more robust 3D imaging performance exploiting its non-uniform intensity distribution and distinct features on potential locations of corresponding point. 
Passive stereo struggles with texture-less scenes and it inevitably requires sufficient ambient illumination. 
Our neural 360$^\circ$ structured light enables robust 3D imaging under low ambient light conditions. 

\subsection{Quantitative evaluation.}
Table~\ref{table:real_comp} provides the quantitative assessment measured for texture-less planes at six different depths: 0.27\,m, 0.5\,m, 0.75\,m, 1.0\,m, 1.7\,m, 2.25\,m.
Neural 360 structured light results in a significantly lower depth reconstruction error in RMSE, up to 5.09$\times$ and 19.6$\times$ less than the multi-dot structured-light\cite{kim2022metasurface} and passive stereo, respectively.
We computed four depth reconstruction error metrics, including inverse mean absolute error (iMAE), inverse root mean square error(iRMSE), mean absolute error(MAE), and root mean squared error(RMSE). 

\subsection{Angle-dependent accuracy.}
We delve further into assessing the 3D imaging performance of neural 360$^\circ$ structured light across the FoV. Our specific focus is on the depth accuracy across five equally-divided sub-FoV angles (36$^\circ$,72$^\circ$,108$^\circ$,144$^\circ$,180$^\circ$). Table \ref{table:real_comp2} presents the depth estimation error for each angle. On average, our neural 360$^\circ$ structured light achieves 7.4$\times$ and 11.4$\times$ lower reconstruction errors compared to both the heuristically-designed multi-dot structured light\cite{kim2022metasurface} and passive stereo methods, respectively.

\begin{table}
\centering
\resizebox{0.6\linewidth}{!}{
    \begin{tabular}{c|c|c|c|c}
    \hline
    Method & iMAE & iRMSE & MAE & RMSE \\
    \hline
    Neural 360$^\circ$ structured light & {0.165} & {0.319} & {0.035} & {0.041} \\
         Multi-dot 360$^\circ$ structured light\cite{kim2022metasurface}  & 0.523 & 0.944 & 0.123 & 0.209  \\
        Passive stereo & 0.633 & 1.08 & 0.652 & 0.807 \\ \hline
    \end{tabular}
} 
    \caption{Quantitative evaluation with depth-reconstruction error. Neural 360$^\circ$ structured light outperforms the heuristically-designed multi-dot structured light\cite{kim2022metasurface} and passive stereo by 5.09$\times$ and 19.68$\times$ in RMSE, respectively.}
  \label{table:real_comp}%
\end{table}

\begin{table}
\centering
\resizebox{0.7\linewidth}{!}{
    \begin{tabular}{c|c|c|c|c|c}
    \hline
    Method & Angle & iMAE & iRMSE & MAE & RMSE \\
       \hline
    \multirow{ 6}{*}{Neural 360$^\circ$ structured light} & 
    0$^\circ$-36$^\circ$ & 0.233 & 0.412 & 0.091 & 0.108 \\
      & 36$^\circ$-72$^\circ$ & 0.286 & 0.462 & 0.097 & 0.109 \\
      & 72$^\circ$-108$^\circ$ & 0.156 & 0.262 & 0.040 & 0.045 \\
      & 108$^\circ$-144$^\circ$ & 0.141 & 0.245 & 0.054 & 0.072 \\
      & 144$^\circ$-180$^\circ$ & 0.160 & 0.251 & 0.077 & 0.085 \\ 
      & Average & {0.195} & {0.326} & {0.072} & {0.084} \\
       \hline
    \multirow{ 6}{*}{Multi-dot 360$^\circ$ structured light\cite{kim2022metasurface}} & 0$^\circ$-36$^\circ$ & 0.527& 1.236 & 0.359 & 0.607 \\
      & 36$^\circ$-72$^\circ$ & 0.390 & 0.563 &	0.174 &	0.194 \\
      & 72$^\circ$-108$^\circ$ & 0.581 & 0.991 & 0.097 & 0.137 \\
      & 108$^\circ$-144$^\circ$ & 0.477 & 0.808 & 0.172 & 0.208 \\
      & 144$^\circ$-180$^\circ$ & 0.622 & 1.082 & 0.290 & 0.339 \\ 
      & Average & 0.528 & 0.936 & 0.218 & 0.297 \\
       \hline
    \multirow{ 6}{*}{Passive stereo} &
    0$^\circ$-36$^\circ$ & 0.745 &	1.378 &	1.229 &	1.707  \\
      & 36$^\circ$-72$^\circ$ & 0.233 &	0.295 &	0.393 &	0.637\\
      & 72$^\circ$-108$^\circ$ & 0.278 &	0.452 &	0.459 &	0.779 \\
      & 108$^\circ$-144$^\circ$ &0.255 &	0.306 &	0.415 &	0.630 \\
      & 144$^\circ$-180$^\circ$ & 0.558 &	0.690 &	0.863 &	1.042 \\ 
      & Average & 0.414 &	0.624 &	0.672 &	0.959 \\
       \hline
    \end{tabular}
}
    \caption{Quantitative evaluation of estimated depth at each FoV group. Neural 360$^\circ$ structured light achieves a lower depth-reconstruction error, up to $7.4\times$ and $11.4\times$ less compared to the multi-dot structured light\cite{kim2022metasurface} and passive stereo, respectively.}
  \label{table:real_comp2}%
\end{table}

\clearpage

\section*{Supplementary Note 11: Additional Experimental Results of 3D Imaging}
\label{sec:qualitative_real}
\begin{figure}[t]
	\centering
		\includegraphics[width=\columnwidth]{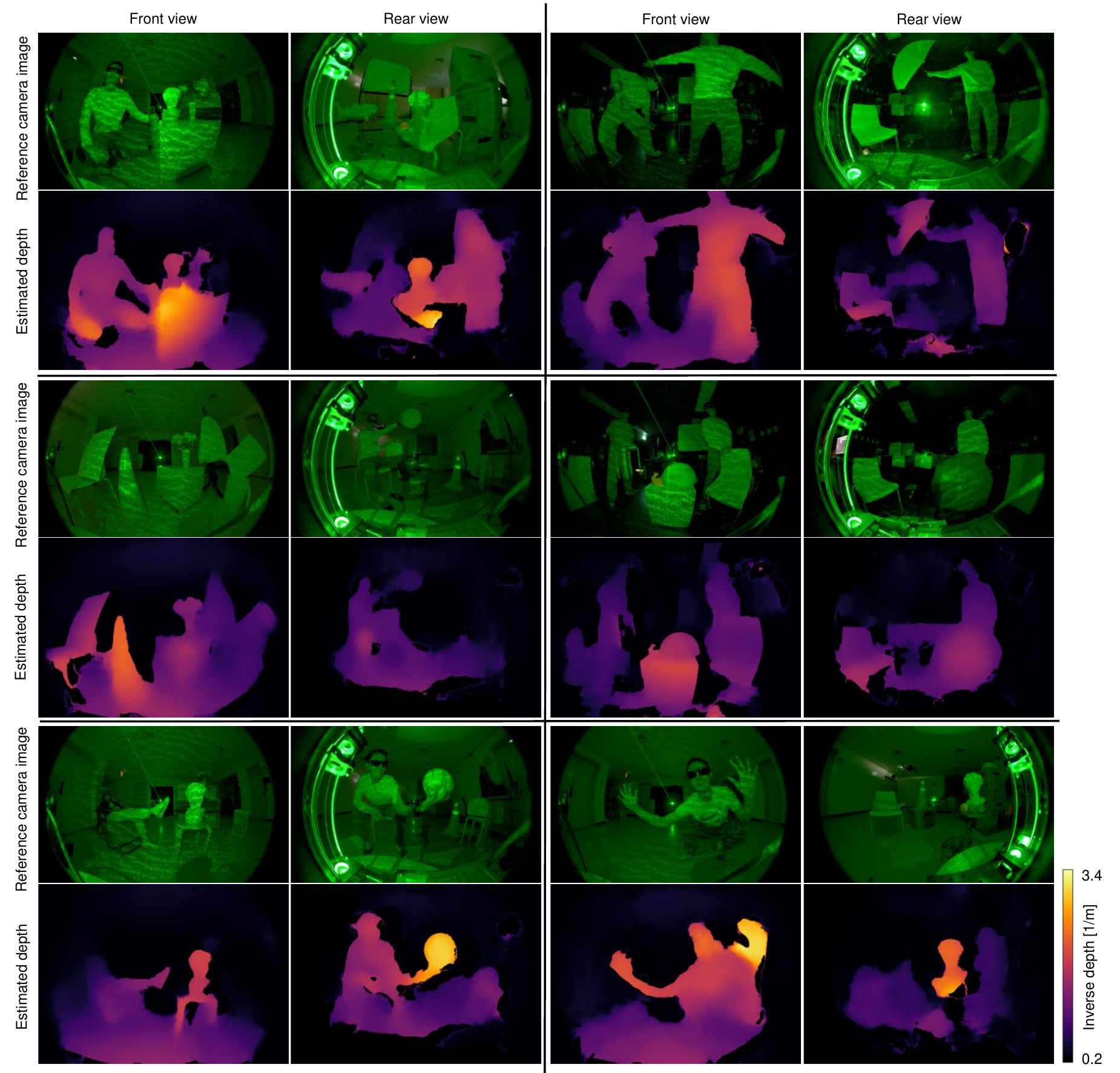}
		\caption{Additional experiment results of 3D imaging with neural 360$^\circ$ structured light. }
		\label{fig:exp_result2}
\end{figure}

This section presents additional qualitative results of various real-world scenes. 
Fig.\ref{fig:exp_result2} shows that neural 360$^\circ$ structured light enables accurate reconstruction on the six additional scenes containing various objects, including furniture, dolls, umbrellas, balls, and human subjects.

\clearpage

\section*{Supplementary Note 12: Additional Synthetic Results of 3D Imaging}
\label{sec:synthetic}
We evaluate the performance of 3D imaging with neural 360$^\circ$ structured light on synthetic scenes. In order to analyze the effectiveness of our neural 360$^\circ$ structured light, comparisons were chosen against different illumination conditions: multi-dot structured light\cite{kim2022metasurface}, random structured light, and passive stereo under sufficient ambient light. The random structured light is produced by a metasurface whose phase is drawn from a uniform random distribution. We trained the same network architectures as depth reconstructors for each case using their respective data. 

\subsection{Quantitative result.} The quantitative results are presented in Table~\ref{table:syn_comp}. Our neural 360$^\circ$ structured light outperforms the heuristically designed multi-dot structured light\cite{kim2022metasurface}, random structured light and passive stereo. Specifically, neural 360$^\circ$ structured light achieves lower reconstruction error, up to 1.77$\times$, 1.7$\times$ and 2.09$\times$ than multi-dot structured light, random structure and passive stereo method, respectively.

\subsection{Qualitative result.} 
\begin{figure}[t]
	\centering
		\includegraphics[width=\columnwidth]{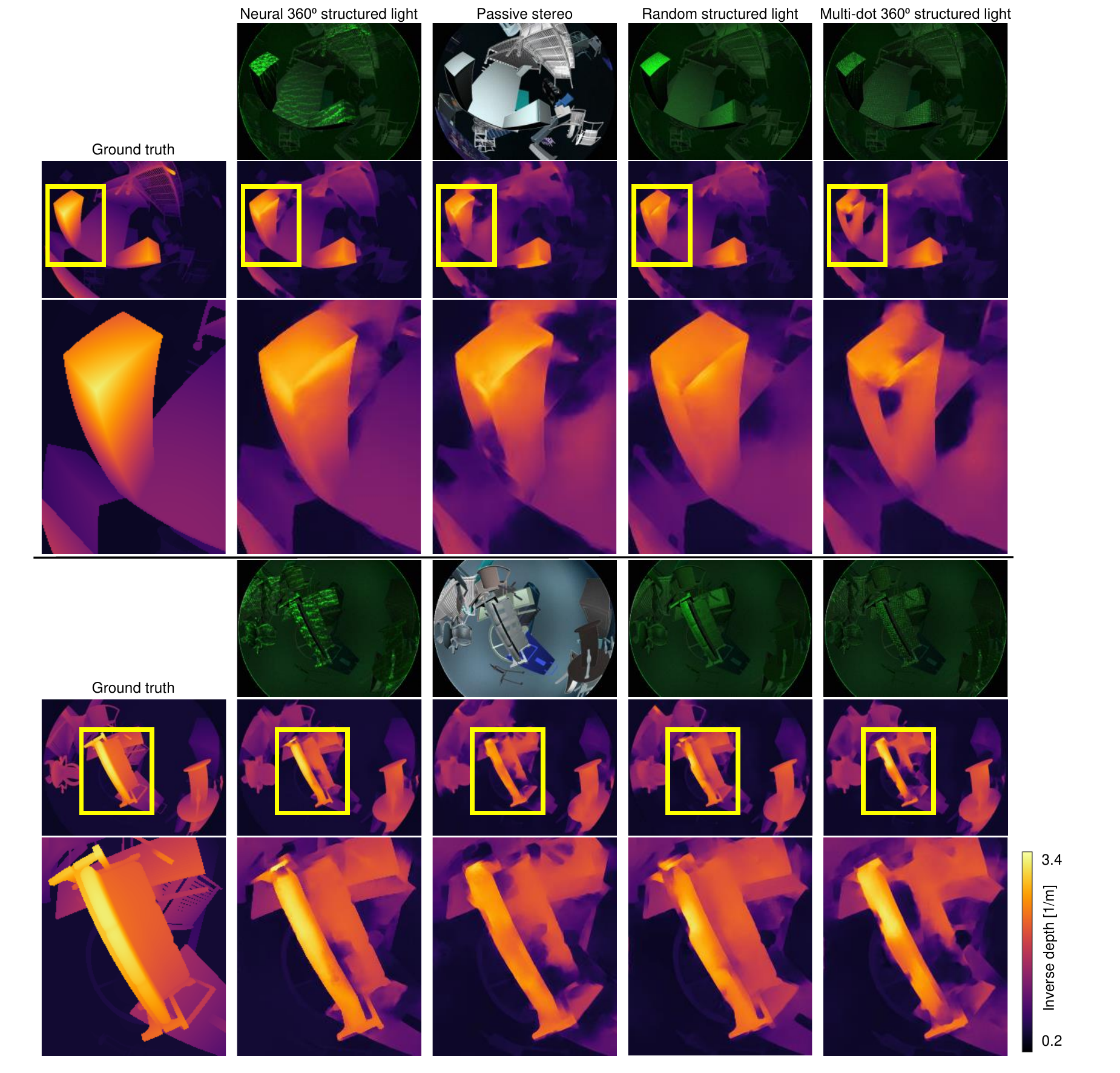}
		\caption{Additional qualitative results on synthetic scenes.}
		\label{fig:synthetic1}
\end{figure}

\begin{figure}[t]
	\centering
		\includegraphics[width=\columnwidth]{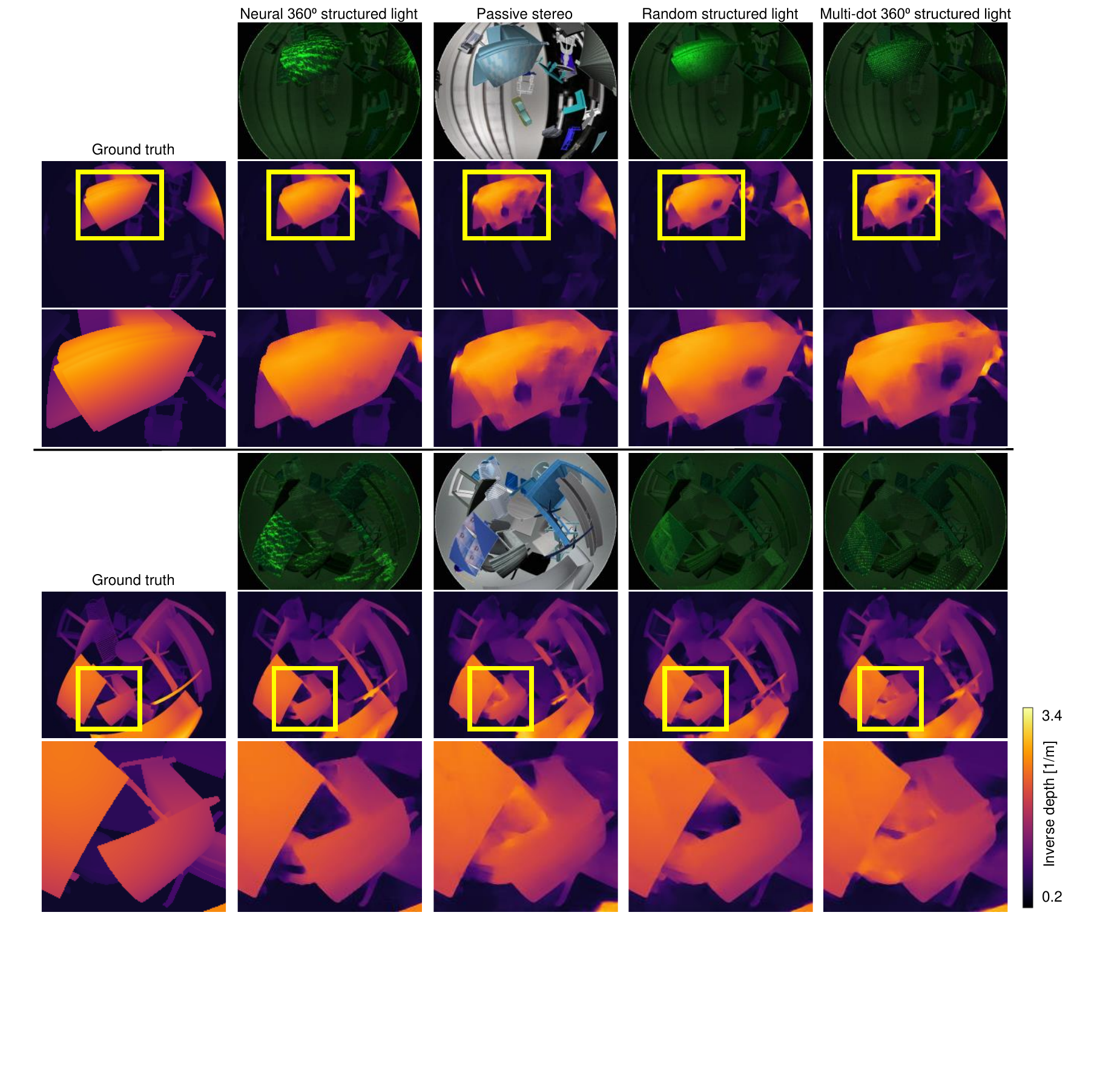}
		\caption{Additional qualitative results on synthetic scenes.}
		\label{fig:synthetic2}
\end{figure}

\begin{figure}[t]
	\centering
		\includegraphics[width=\columnwidth]{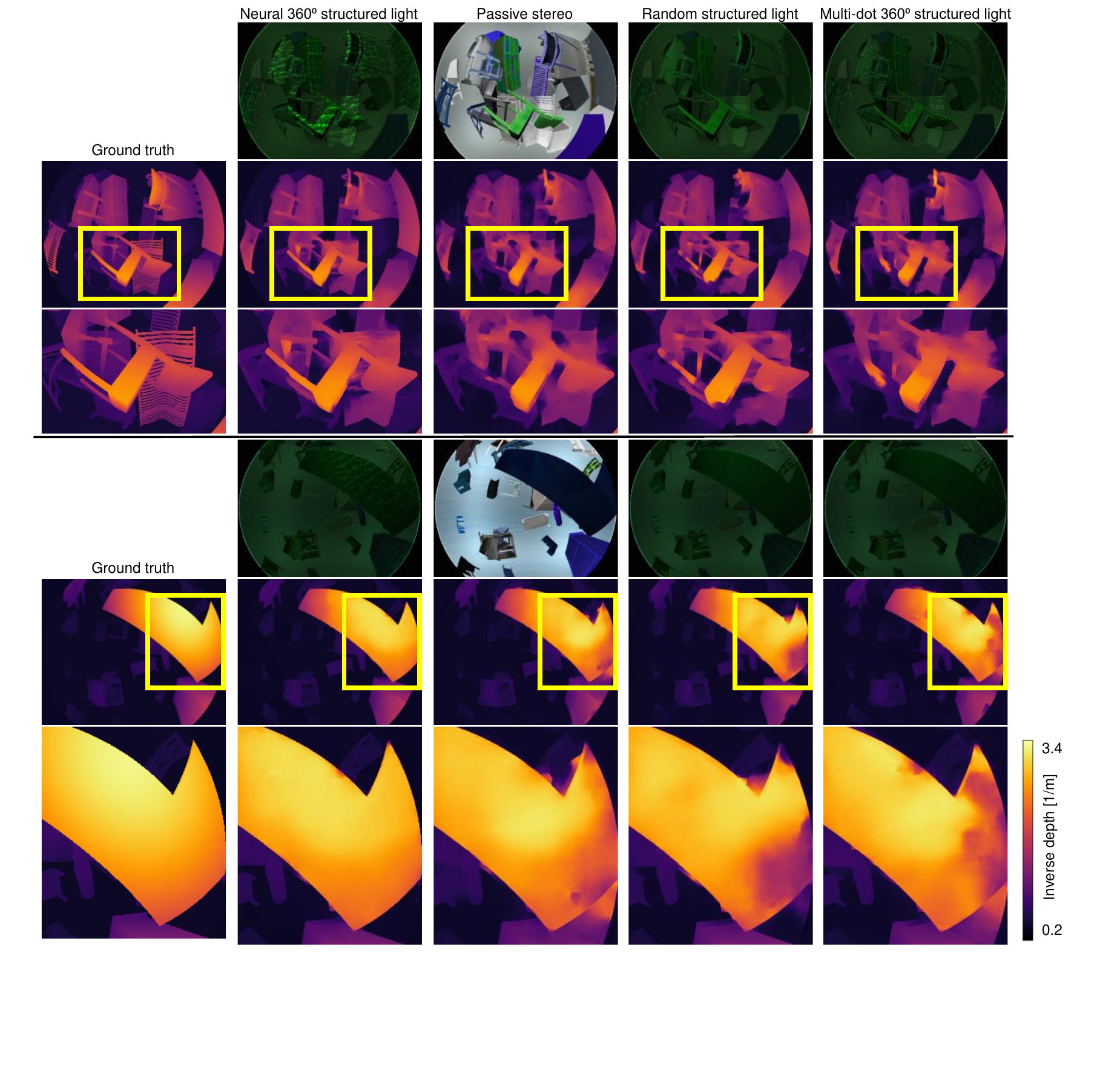}
		\caption{Additional qualitative results on synthetic scenes.}
		\label{fig:synthetic3}
\end{figure}

\begin{figure}[t]
	\centering
		\includegraphics[width=\columnwidth]{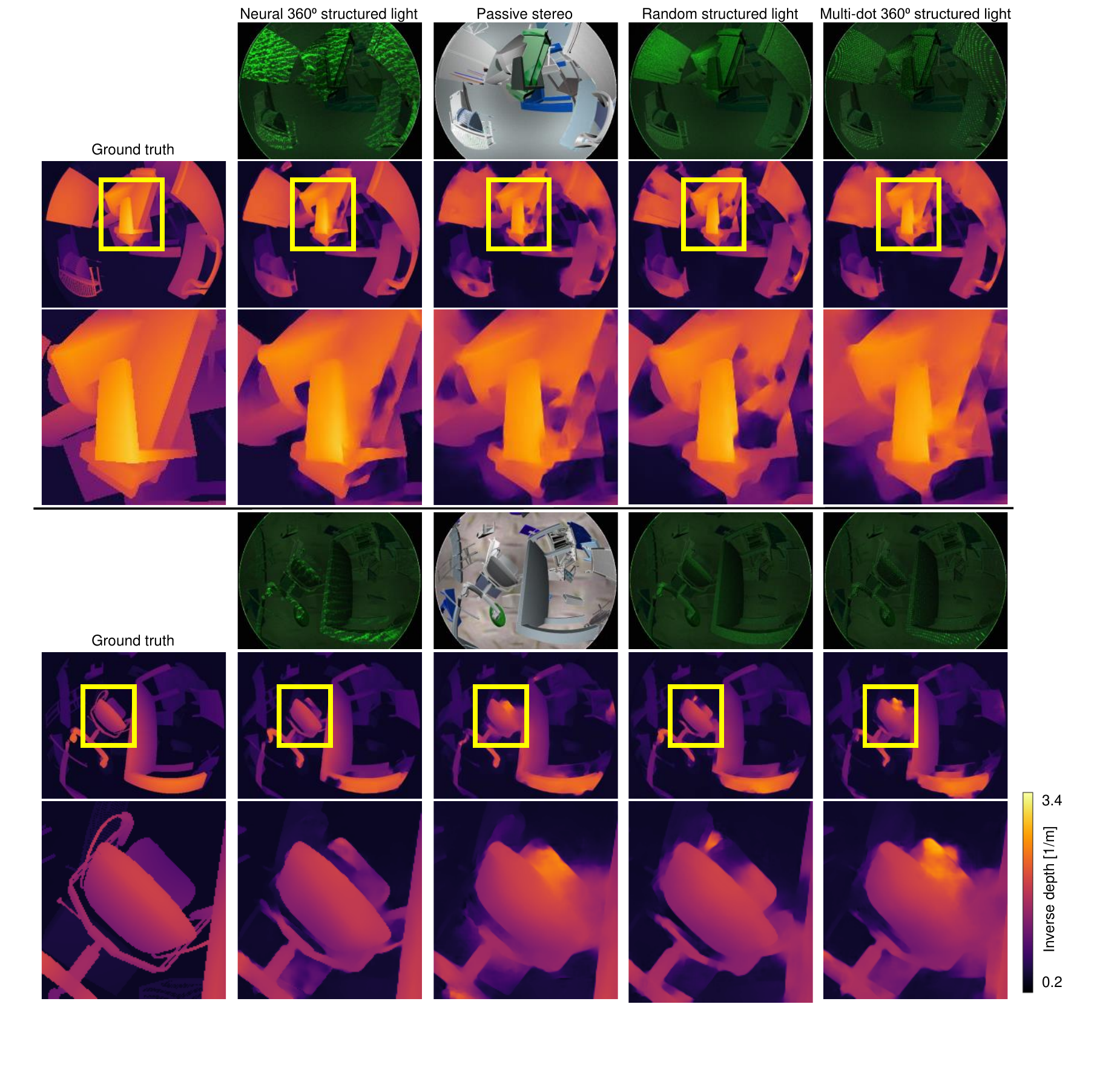}
		\caption{Additional qualitative results on synthetic scenes.}
		\label{fig:synthetic4}
\end{figure}

The qualitative results are illustrated in Fig.~\ref{fig:synthetic1},~\ref{fig:synthetic2},~\ref{fig:synthetic3},~\ref{fig:synthetic4}, which include rendered images and estimated depth for each comparative method. In the qualitative evaluation, our neural 360$^\circ$ structured light outperforms the other methods. Notably, in texture-less scenes, the performance gap is more pronounced. In Fig.~\ref{fig:synthetic3}, regarding the texture-less square-shaped object in the lower row, other methods fail to estimate the correct depth while our method successfully reconstructs the depth information.

\begin{table}
\centering
\resizebox{0.6\linewidth}{!}{
    \begin{tabular}{c|c|c|c|c}
    \hline
    Method & iMAE & iRMSE & MAE & RMSE \\
    \hline
        Neural 360$^\circ$ structured light & {0.072} & {0.202} & {0.237} & {0.547} \\
         Multi-dot 360$^\circ$ structured light\cite{kim2022metasurface}  & 0.121 & 0.273 & 0.421 & 1.021  \\
         Random structured light & {0.116} & {0.261} & {0.403} & {0.717} \\
        Passive stereo & 0.149 & 0.321 & 0.492 & 0.859 \\ \hline
    \end{tabular}
} 
    \caption{Quantitative evaluation on synthetic scenes. Neural 360$^\circ$ structured light outperforms the heuristically-designed multi-dot structured light\cite{kim2022metasurface}, random structured light, and passive stereo.}
  \label{table:syn_comp}%
\end{table}

\clearpage

\bibliographystyle{naturemag}
\bibliography{reference}